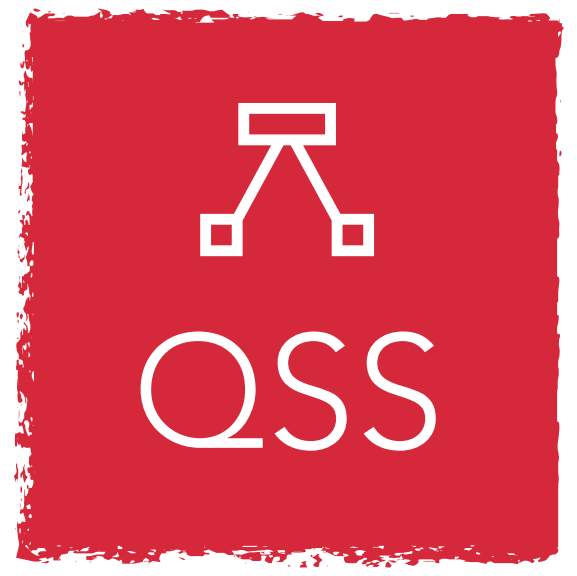

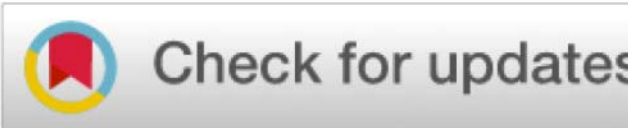

**Corresponding Author:**
Marek Kwiek
kwiekm@amu.edu.pl

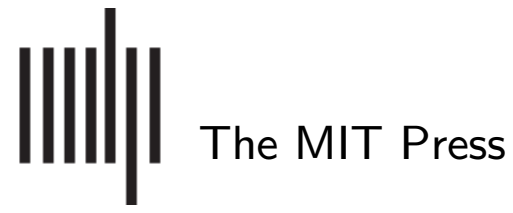

RESEARCH ARTICLE

# Young male and female scientists: A quantitative exploratory study of the changing demographics of the global scientific workforce

**Marek Kwiek**[1,2] and **Lukasz Szymula**[1,3]

[1]Uniwersytet im Adama Mickiewicza w Poznaniu, Poznan, Poland
[2]DZHW GmbH, Hannover, Germany
[3]University of Colorado, Boulder, CO, USA

## ABSTRACT

In this study, the global scientific workforce is explored through large-scale, generational, cross-sectional, and longitudinal approaches. We examine 4.3 million nonoccasional scientists from 38 OECD countries publishing in 1990–2021. Our interest is in the changing distribution of young male and female scientists over time across 16 science, technology, engineering, mathematics, medicine (STEMM) disciplines. We unpack the details of the changing scientific workforce using age groups. Some disciplines are already numerically dominated by women, and the change is fast in some and slow in others. In one-third of disciplines, there are already more youngest female than male scientists. Across all disciplines combined, the majority of women are young women. And more than half of female scientists (55.02%) are located in medicine. The usefulness of global bibliometric data sources in analyzing the scientific workforce along gender, age, discipline, and time is tested. Traditional aggregated data about scientists in general hide a nuanced picture of the changing gender dynamics within and across disciplines and age groups. The limitations of bibliometric data sets are explored, and global studies are compared with national-level studies. The methodological choices and their implications are shown, and new opportunities for how to study scientists globally are discussed.

## 1. INTRODUCTION

We explore the changing demographics of the global scientific workforce from the combined perspectives of age, gender, academic discipline, and time. Our approach is large scale, generational, and both cross-sectional and longitudinal. With this approach, we examine 4.3 million nonoccasional scientists (defined as scientists with an output of at least three Scopus-indexed articles) publishing over the past three decades (1990–2021). Our longitudinal interest is in the changing distribution of male and female scientists over time across different academic age groups—especially young scientists with no more than 10 years of publishing experience—across 16 science, technology, engineering, mathematics, medicine (STEMM) disciplines. The present research focuses on the OECD area: Whenever the term “global” is used in our results, technically, it refers to “38 OECD countries.”

Large-scale and longitudinal approaches toward studying the differences in academic careers by gender, age, and discipline have been used only recently, accompanied by increasing access to a digital national and global, commercial and noncommercial workforce, and administrative, and bibliometric databases, such as Web of Science, Scopus, and Microsoft Academic Graph (MAG), as well as Academic Analytics and Digital Bibliography and Library Project (DBLP) for the United States and CRISTIN for Norway or POL-on for Poland (Boekhout, van der Weijden, & Waltman, 2021; Elsevier, 2020; King, Bergstrom et al., 2017; Kwiek & Roszka, 2021b; Larivière, Ni et al., 2013; Nielsen & Andersen, 2021; Nygaard, Piro, & Aksnes, 2022; Robinson-Garcia, Costas et al., 2020; Savage & Olejniczak, 2021; Way, Morgan et al., 2017; Zhang, Wapman et al., 2022). Advances in author disambiguation methods allow us to examine academic careers at a global scale and over time. The global scientific workforce can be studied from new perspectives: Our focus on young male and female scientists is one such perspective. A generational approach to the changing scientific workforce—especially age distribution by specific age groups by gender (e.g., the young vs. the old)—has not been applied at a global level.

Currently, the participation of men and women in science can be studied over time with a previously unattainable level of detail across countries, institutions, and disciplines, as well as across age and seniority groups. Publications and their authors can be examined through temporal, topical, geographic, and network analyses or connected to time, themes, places, and other scientists (Börner, 2010, pp. 62–63). In our study, we have followed three previous studies: Huang, Gates et al. (2020), who reconstructed the complete publication history of over 1.5 million scientists to examine gender inequality in scientific careers globally (83 countries, 13 disciplines), Boekhout et al. (2021), who traced the publication careers of about 6 million male and female scientists in 1996–2018, and King et al. (2017), who examined 1.5 million research papers from the JSTOR bibliometric database to show gender differences in self-citation rates across disciplines and time. However, none of these studies focused on young male and female scientists.

### 1.1. Academic Careers: Global Studies

Huang et al. (2020), Boekhout et al. (2021), and King et al. (2017) focused on the various aspects of scientific careers globally, used large bibliometric data sets and methods, and clearly applied the male/female distinction; however, age categories were not combined with gender. Consequently, although gender, academic age, and discipline variables were used, the changing demographics of the global scientific workforce over time were not examined; specifically, the participation of young male and female scientists was not the focus. In this research, we present a comprehensive overview of the participation in science of male and female scientists changing over the past three decades.

Although Huang et al. (2020) used Web of Science as their source database (with Microsoft Academic Graph and DBLP as supplementary data sets), Boekhout et al. (2021) used Scopus, which offers more complete data on first names, which is essential for gender inference. Both papers did not refer to researchers whose gender was unknown in their analyses and included authors with at least three publications. King et al. (2017) used the scholarly database JSTOR and used the idea of authorships as unique author–paper pairs (King et al., 2017, p. 4).

In a longitudinal study, gender differences in publishing career lengths and dropout rates were studied in Huang et al. (2020); they used a career length matching design to study the relationship between career length and total productivity (412,778 female authors were matched with 412,778 male authors). A large proportion of the observed gender gaps were rooted in gender-specific dropout rates and subsequent gender gaps in publishing career length and total productivity (Huang et al., 2020, p. 4615).

The authors reconstructed the complete publication histories of all gender-identified authors from Web of Science whose publishing careers ended between 1955 and 2010. Their focus was on career-wise gender differences in productivity and impact. The gender gap was found to be increasing over time and persistent. Each year, female scientists had a 19.5% higher risk to leave academia compared with male scientists—which is a major cumulative advantage for male authors over time (Huang et al., 2020, p. 4613).

In their longitudinal research design, Boekhout et al. (2021) examined publication productivity for men and women who started their careers as publishing researchers in 2000, 2005, and 2010, using full counting and fractional counting approaches. They showed an increasing trend in the percentage of women starting their careers as publishing researchers, from 33% in 2000 to 40% in 2015. Instead of considering entire publication careers (as in Huang et al., 2020), the authors compared the productivity of male and female scientists in specific years in their careers, showing that male scientists have consistently higher publication productivity than female scientists, regardless of the year in which they started their career and period in their career, with differences in the range of 20–35% (full counting) and 25–40% (fractional counting) in favor of male productivity (Boekhout et al. (2021, p. 9); see Kwiek (2016) and Kwiek (2018) on research top performers).

Finally, gender gap in self-citations across time and disciplines was examined in King et al. (2017), with men in the past few decades self-citing 70% more than women. Women were also more than 10 percentage points more likely not to cite their previous work at all. The authors linked self-citations to larger themes of inequality in science and cumulative advantage in science careers (because self-citations increase citations). They reported the gender self-citation gap to be stable over the past 50 years. Compared with men, women have been overrepresented in the zero self-citation category and underrepresented in terms of citing their papers (King et al. (2017, p. 8); for a general overview of gender differences in science, see Halevi (2019) and Sugimoto and Larivière (2023)).

Other examples of recent influential large-scale studies of academic careers and global publishing, collaboration, and impact patterns include Robinson-Garcia et al. (2020), who studied gender differences in archetype career tasks, Larivière et al. (2013), who examined global gender disparities in science, Nielsen and Andersen (2021), who studied the global citation elite, and Ioannidis, Boyack, and Klavans (2014), who focused on the continuously publishing core in global science. Robinson-Garcia et al. (2020) examined 71,000 publications from PLOS journals, with 350,000 distinct authors, to profile scientists across three task specializations and the changes in their career stages. They used four career stages (junior, early career, midcareer, and late career, using the years elapsed from first publication); and three archetype tasks were studied: leader, specialized, and supporting. Scientists were reported to be unevenly distributed by gender in each archetype, with men being more likely to be leaders and women representing the specialized archetype in early career stages, which is the most important for later academic promotions (Robinson-Garcia et al., 2020, p. 12).

The authors constructed publication histories and grouped publications by career stages, using the minimum threshold of five publications, academic age based on the first publication, and the 90% accuracy threshold in assigning gender to individual scientists.

Global gender disparities in science were also studied in Larivière et al. (2013). The authors used 5.5 million papers and 27.3 million authorships to show that, globally, women account for fewer than 30% of fractionalized authorships and are similarly underrepresented regarding first authorships. Female collaborations tend to be more domestically oriented than collaborations of men from the same country, and when a woman is in a prominent author position (sole authorship, first authorship, and last authorship), a paper attracts fewer citations than when a man is in one of these roles (Larivière et al., 2013, p. 213; see also Kwiek, 2020). Based on a data set of 4 million authors and 26 million papers, Nielsen and Andersen (2021) studied the rise in global citation inequality, with a small stratum of elite scientists accruing increasing citation shares. They examined the temporal trends in the concentration of citations at the author level, focusing on differences in the degree of concentration across fields, countries, and institutions. They found that the top 1% most cited scientists ("the citation elite") have increased their cumulative citation shares from 14% to 21% between 2000 and 2015 without increasing their general productivity level (in fractional counts) or impact per paper. The authors in the citation elite increasingly reside in Western Europe and Australasia, with a decreasing share of top-cited scientists in the United States (Nielsen & Andersen, 2021, p. 4).

Nielsen and Andersen (2021) and Ioannidis et al. (2014), in contrast to Larivière et al. (2013), did not disaggregate their results by gender or by academic age. However, Nielsen and Andersen (2021) noted that citation-elite membership is strongly correlated with age and suggested future research within and across age cohorts.

Finally, in their study of the "continuously publishing core" of the global scientific workforce, which was based on 15.2 million publishing scientists from 1996 to 2011, Ioannidis et al. (2014) showed that less than 1% of scientists—or about 150,000—published their research each year in the studied 16-year period, accounting for as much as 87.1% of papers with more than 1,000 citations. The authors examined what they termed "uninterrupted continuous presence" (UCP) in the Scopus-indexed literature, analyzing who maintains their presence each and every year for many years, which is another dimension of the "elite" or "core" status in science. The proportion of scientists with a UCP presence is very limited, but they account for the lion's share of researchers with a high citation impact.

As in the case of our present research, the authors used Scopus author identifiers rather than attempting to disambiguate authors on their own. The UCP-birth and UCP-death years of an author were the calendar years that start and end their chain of uninterrupted, continuous, and annual publications (Ioannidis et al., 2014, p. 2). The 1% of scientists was found to be a very influential core of science, with much higher citation metrics than other researchers. Although the global scientific workforce is enormous, its continuously publishing core is very limited, with many departments or institutions having none or very few researchers who belong to this group (Ioannidis et al., 2014, p. 9).

The analysis did not consider variables such as gender or academic age, without disaggregating the data into countries, men and women, or career stages, with the assumption being that the UCP presence, by definition, refers to older age cohorts and higher seniority levels.

### 1.2. Academic Careers: Examples from the United States

Also, large-scale, national-level studies of academic careers in the United States have been increasingly precise in terms of gender, discipline, and age determination. For instance, Way et al. (2017) examined the traditional "rapid rise, gradual decline" narrative about productivity patterns, showing that this pattern holds for only 20% of individual faculty (whereas for the remaining 80%, there is a rich diversity of patterns). Using a DBLP data set of 200,000 publications and career trajectories of 2,453 tenure-track faculty from computer science departments and their CV data, the authors showed how much diversity is hidden behind average academic career trajectories, creating inaccurate pictures of productivity patterns. The authors examined the productivity trajectories of individual researchers in an entire field of research and showed that 60 years of research on aggregate trends needs a revision in view of the conclusions derived from studies based on much larger and more comprehensive data sets.

Although academic experience was heavily used, gender differences were not studied. (Similarly, using the Academic Analytics commercial database, Savage and Olejniczak (2021) showed that the career publication activity of U.S. scientists does not follow the traditional "peak-and-decline" pattern described in earlier studies.)

Finally, using a combination of data sources, such as Academic Analytics, Web of Science, and the NSF Survey of Graduate Students and Postdoctorates in Science and Engineering, Zhang et al. (2022) showed that the disproportionate productivity of scientists in U.S. elite institutions can be largely explained by their substantial labor advantage: their better access to externally funded graduate and postdoctoral labor.

They used a matched pair design in which one midcareer researcher in the pair moved to a working environment with more available labor, and the other moved to an environment with less available labor ($n$ = 778 faculty), with detailed productivity data for 78,000 faculty across 25 scientific disciplines. The association of institutional prestige with greater productivity was explained by greater available funded labor, which drove larger group sizes, thereby increasing group productivity (Zhang et al., 2022, p. 6).

The productivity dominance of researchers at elite institutions was found not to result from inherent characteristics (such as differences in talent) but rather can be explained by the greater labor resources provided to them in more prestigious environments. The authors showed the pivotal role of funded labor and external research funds in explaining the dominance of elite institutions but did not distinguish between academic careers by male and female scientists.

#### 1.3. Academic Careers: Cross-National Survey-Based Studies

Additionally, recent changes in academic careers have been widely documented in a separate line of research: the literature generated by cross-national comparative survey designs. Large-scale comparative studies have included books on the United States (Cummings & Finkelstein, 2012) and Japan (Arimoto, Cummings et al., 2015), as well as Europe (Kwiek, 2019), with a focus on academic work (Fumasoli, Goastellec, & Kehm, 2015); recruiting and managing the academic profession (Teichler & Cummings, 2015); internationalization of teaching and research (Huang, Finkelstein, & Rostan, 2014); the relevance or impact of research (Cummings & Teichler, 2015); and the various faces of internationalization of the academic profession (Calikoglu, Jones, & Kim, 2023).

Cross-national comparative studies from this line of research (summarized in Carvalho, 2017) have provided excellent complementary sources to studies of bibliometric data sets: they are relatively small scale, with national data sets usually in the range of 1,000 to 4,000 observations, and focusing on issues not obtainable through bibliometric data (e.g., personal opinions, perceptions, and feelings; family life and motherhood in academia; university governance and management; job satisfaction).

Although young men and women are often examined in survey-focused literature under the label of juniors (contrasted with seniors), the number of cases is usually too limited to analyze the differences by disciplines, and research designs only allow for cross-sectional analyses.

### 1.4. Academic Careers: Statistical Reports

There have also been several reports on women in science over the past few years, with different geographical focus (see, e.g., NSF [2023] on the United States; EC [2021] on the European Union; and globally, Elsevier [2015, 2017, 2020]). Specifically, the Elsevier reports on gender differences in research provide statistics and analyses on similar topics to ours.

There are, however, important differences between our approach and those in the reports in the study design, research focus, methodology, and results.

Most importantly, our focus is on specifically defined (via age groups based on academic age or academic experience) young male and female scientists and their changing participation across STEMM (traditional STEM disciplines plus medicine) and over time in 38 OECD countries; we have used a combination of horizontal (men compared with women in the same age group and across time) and vertical (men and women separately disaggregated into age groups and compared across time) approaches; and our unit of analysis is the individual scientist with specific characteristics derived from large-scale bibliometric data sets, especially nonoccasional status in science, which requires meeting the threshold of having at least three research articles published.

The first report is a cross-national comparative study of European countries. *She Figures 2021* (EC, 2021), which covered 44 countries, used the data extracted from Eurostat statistics on education, research, and development; professional earnings and human resources in science and technology; and the Scopus database. The report discussed the labor market participation of researchers, working conditions of researchers, career advancement and participation in decision-making, and research and innovation output, all of which is outside of the scope of our paper. Interestingly, the report provided examples of actions taken to promote gender balance in science across different countries (e.g., EC, 2021, pp. 183–185). It provided the data on women among doctoral graduates across broad fields of study, including the STEM fields, with a general conclusion that women remained underrepresented in most STEM fields, with little or no progress since 2015 (EC, 2021, p. 39). The report discussed the changes in the Glass Ceiling Index (GCI), a relative index comparing the proportion of women in academia to the proportion of women in top academic positions for 2015–2018, with the GCI decreasing in most countries studied (EC, 2021, pp. 192–194).

The report provided an analysis of the gender gap among active authors, who were defined as those who produced 10 or more papers over the past 20 years (2000–2019) and at least one paper in the past 5 years or those who produced four or more papers in the past 5 years. The report used three seniority levels estimated via the time elapsed since an author's first publication in Scopus (early-stage, middle-stage, and senior authors) and the ratios of women to men among active authors by broad fields and countries.

The major takeaway from the report is that, among early-stage authors, the gender gap was smaller, but as seniority level increased, the gender gap widened to twice as many male as female authors. Women were the least represented in the natural sciences and engineering and technology and most represented in medical and health sciences and agricultural and veterinary sciences (EC, 2021, p. 218).

The second report is a single-nation study of the United States. The NSF report (NSF, 2023) analyzed women, minorities, and persons with disabilities in the STEM workforce in the United States, specifically using gender, race and ethnicity, and degree levels. This was a single-nation report with no references to academic publications or career stage or academic age combined with gender. The report used the notion of the STEM workforce as defined in labor force statistics, which included workers in science and engineering (S&E) and S&E-related and middle-skill occupations. In 2021, nearly a quarter (24%) of individuals in the U.S. workforce were employed in STEM occupations (NSF, 2023, p. 8).

The three reports from Elsevier came the closest to our study in terms of their methodologies because they used the Scopus data set and individual Scopus identifiers to define individual scientists. The single-nation report on Germany (Elsevier, 2015) paved the way for the report on 12 geographies (Elsevier, 2017) and a more comprehensive report on gender in research in 16 geographies (Elsevier, 2020). The first report linked Germany's relatively low share of female researchers among European countries to its research focus on physical sciences and mathematics, which are traditionally male-dominated fields. Female researchers in Germany are reported to be concentrated in medicine (and social sciences, which is not discussed in our paper).

Consistent with the findings from other studies, the share of women was lower among senior researchers than among junior researchers, with a "leaky" pipeline in science careers: A higher proportion of women than men moved out of the world of science while moving up the academic career ladder (Elsevier, 2015, p. 9). In the report, the research productivity and citation impact of men and women per year by seniority level was compared. Across the three seniority categories, male scientists had higher productivity compared with female scientists; however, gender gaps in citation impact were visible mainly for junior and middle-senior levels and almost disappeared for senior levels (Elsevier, 2015, pp. 12–16).

Another Elsevier report (Elsevier, 2017) provided detailed methodological and data sources appendixes, with major procedures explained and definitions provided. Specifically, name and gender disambiguation for researchers were described, as were the concepts of "active researchers," "authors," "inflow," "outflow," "migratory," "transitory," and "nonmigratory" researchers (Elsevier, 2017, pp. 84–87). In the report, the proportion of men and women among researchers in 12 comparator countries and regions in the two time periods (1996–2000 and 2011–2015) was analyzed by subject areas for each gender and comparator.

The key findings were that the proportion of women among researchers has increased in all comparators and that women tend to specialize in biomedical fields and men in physical sciences (Elsevier, 2017, p. 19). The report did not refer to career stages and gender, especially not analyzing the participation of young female scientists in science.

Finally, the most detailed report on men and women in science today was again by Elsevier (2020). In its comprehensive approach, it provided the gender-disaggregated data on science participation, publishing career and mobility, and collaboration networks across 15 countries and the EU-28, using the Scopus database. The analyses used four broad subject clusters (physical sciences, health sciences, life sciences, and social sciences) and 27 major subject areas (e.g., mathematics, medicine, biochemistry).

The major procedures were described in detail in the appendixes: author definition and disambiguation (Scopus Author Profiles), active authors, author country and subject area assignation, country selection, author gender inference, author publication history, author mobility, and author collaboration network analysis (Elsevier, 2020, pp. 119–133). "Active authors" were defined as those who authored at least two publications in the study periods (1999–2003 and 2014–2018).

The report showed that men are more highly represented among authors with a long publication history and women with a short publication history (Elsevier, 2020, p. 37). In terms of publication output, on average, women published less than men in a 5-year period in every country assessed, regardless of authorship position (Elsevier, 2020, pp. 37–38) and in terms of citation practice, the average Field-Weighted Citation Impact of men was higher than that of women (Elsevier, 2020, p. 41).

In the report, the concept of "academic age" was not used, and researchers' academic stages were not defined accordingly. However, four groups based on the length of publication history were used. The report did not focus on the participation of young women (and men) in science across disciplines and over time, but it provided excellent methods to study academic careers using bibliometric data sources.

Our paper examines what we can know—based on available global data sources of the bibliometric type—about the changing demographics of the scientific workforce globally and over time. We wanted to explore how useful the potential global data sources can be in analyzing the scientific workforce along the combined four dimensions of gender, age, discipline, and time. We tested how demographic transformations of the global science profession can be measured using new data sources, hence transgressing the traditional approach in which national statistics from national statistical offices are aggregated, as in the OECD, UNESCO, and European Union scientific workforce data sets.

In the present research, we contribute to the discussion of the advantages and disadvantages of using global publication and citation databases—or "structured" Big Data (Holmes, 2017; Salganik, 2018; Selwyn, 2019)—in global academic profession studies in which the data on gender, age, and disciplines have traditionally been available almost exclusively cross-sectionally (single points in time), mostly on a small national scale (through case studies) and increasingly on a small international comparative scale through cross-national survey research of the academic profession. We unpack the details of the changing scientific workforce using 10 5-year age groups within each discipline from a longitudinal perspective.

### 1.5. Women in STEM: Theoretical Background

The global picture of young men and women in science is a general overview of their representation across disciplines around the world. This global picture shows patterns and trends over time and across disciplines. The representation varies widely at the national level because of social, economic, political, and cultural factors. There are countries with stronger policies and initiatives in place to encourage women to pursue STEM education, with a larger pool of female graduates entering doctoral programs and the academic profession; and there are countries where cultural and societal attitudes may discourage women from pursuing careers in science.

As a result, although variations by country can be huge, our interest is in global cross-disciplinary differences changing over time. Targeted interventions and policies to address the underrepresentation of women in some disciplines, here resulting from both low entering shares and high exiting shares for women, young and older alike, need to be developed at a national level.

By examining the national picture, we can obtain a more nuanced understanding of the representation of women in science, leading to more effective strategies at the level of disciplines. In the present research, we do not consider career breaks, which may be more common among women because of caregiving responsibilities, and we do not consider the broader context of gender and work–family balance.

Young scientists—young female scientists in particular—face unique challenges and barriers to enter, continue, and advance in science careers. Apart from underrepresentation of women in science, there are implicit biases (stereotyping and discrimination against women in STEM); unwelcoming or hostile workplace cultures, especially in male-dominated disciplines; and challenges related to work–life balance and motherhood responsibilities, possibly leading to career interruptions and slower career progression. As the Elsevier (2020) report showed, women continue to face significant challenges at every stage of their careers: they are underrepresented in senior positions, less likely to collaborate internationally, more likely to experience career breaks, less likely than men to publish articles in high-impact journals, and cited less frequently, on average (see Dusdal & Powell, 2021; Kwiek & Roszka, 2021a, 2022b; Sugimoto & Larivière, 2023; Tang & Horta, 2023).

Although both men and women leave science in some proportions, the attrition for women in STEM is higher. Major theories about women leaving science are the "leaky pipeline" theory, the "chilly climate" hypothesis, and the "self-selection" hypothesis: Leaky pipeline theory suggests that there is a significant loss of talent at every stage of the academic career pipeline, from female graduates to female postdocs to female assistant professors and to female tenured professors because of systemic barriers such as bias and discrimination (see, e.g., Sexton, Hocking et al., 2012; Shaw & Stanton, 2012; Sheltzer & Smith, 2014; Wolfinger, Mason, & Goulden, 2008); chilly climate theory suggests that a hostile or unwelcoming work environment in STEM disciplines can discourage women from pursuing careers (see, e.g., Cornelius, Constantinople, & Gray, 1988; Hall & Sandler, 1982; Maranto & Griffin, 2011; Morris & Daniel, 2008); and self-selection theory suggests that women are underrepresented in STEM disciplines because they are less interested in these disciplines because of societal and cultural factors that discourage them (see, e.g., Britton, 2017; Hyde, Fennema et al., 1990; Whitt, Nora et al., 1999).

Finally, the glass ceiling metaphor is used to describe gender inequality in science from a different angle: an invisible barrier that prevents women from advancing to higher levels of leadership and power within organizations, including universities. There are systemic barriers that make women unable to reach the opportunities and rewards above them. An invisible barrier limits professional recognition, with few women becoming full professors (see e.g., Morrison, White, & Van Velsor, 1987; Tang, 1997).

### 1.6. Research Questions

We focus on the individual scientists (with their unique identity) as the unit of analysis, rather than publications. Although a bibliometric data source is used (Scopus raw data provided to us by Elsevier's International Center for the Studies of Research [ICSR] Lab through a multiyear collaboration agreement), our focus is on scientists and their attributes rather than publications and their properties. Our microdata show gender, academic age or academic experience, discipline, country, and publications and their types (lifetime); we turn bibliometric data sources on publications into data sources on individuals.

Our three research questions regarding publishing and nonoccasional STEMM scientists are as follows:

1. What is the global disciplinary distribution of young male and female scientists?
2. How do the global gender and age distributions of scientists across disciplines change over time, especially for young male and female scientists?
3. How is the participation in science of female scientists changing over time and across disciplines, and what are the disciplinary gender participation trends?

## 2. DATA AND METHODS

### 2.1. Data

The major characteristics of the longitudinal study population for 1990–2021 (4,314,666 scientists, including 1,645,860, or 38.15%, female) are presented in Table 1. The major characteristics of the cross-sectional study subpopulation for 2021 (1,502,792 scientists, including 579,399, or 38.55% female) are presented in Table 2. Our population was constructed as follows (we refer to the population rather than the sample because we have all scientists, with their attributes, as units of analysis): First, to determine the number of scientists, unique authors of publications (type: journal article, conference paper in a book, or a journal) who published their works in 1990–2021 were selected. For this selected group of authors, the years of their research activities were determined. The resulting set of scientists was then narrowed down according to a package of five restrictions: (a) an OECD country; (b) a STEMM discipline; (c) gender (binary approach: man or woman); (d) a nonoccasional status in science: a minimum scientific output defined as three publications throughout the scientist's career (lifetime); and (e) academic age, or the time passed since the first publication, here in the 1–50 years range.

The minimum output in lifetime publication history allowed us to limit our population to nonoccasional scientists, that is, scientists functioning in the scientific community more than accidentally. Additionally, scientists with one or two publications in the Scopus database are more likely to result from mistakes made by author name disambiguation algorithms (see Boekhout et al., 2021, p. 3). Generally, in terms of author name disambiguation, Scopus is more accurate than Web of Science (Sugimoto & Larivière, 2018, p. 36). Then, for each scientist, academic experience in full years, beginning in the year of the first publication of any type, was determined. For each year of a scientist's research activities, the length of their academic experience and membership in the corresponding academic age group were determined. We used a population for 1990–2021 for longitudinal analyses, a subpopulation for 2021 for a cross-sectional analysis, and the two subpopulations for 2000 and 2021 for analyses comparing two points in time. Figure 1 summarizes the population's design.

### 2.2. Methods

In this section, we present the five basic procedures to unambiguously define the attributes of the scientists in our population. We initially used raw data for 2020 and before, here based on the Scopus database version dated 18 August 2021. The raw data were made available to us by Elsevier under an agreement with the ICSR Lab. Finally, the Scopus database version for 2021 and before, dated October 21, 2022, was used.

To obtain the results at the aggregate level, the operation in the ICSR Lab relied on the use of the Databricks environment, which allowed for managing and executing cloud computing with Amazon EC2 services. The scripts to generate the results were written using the PySparkSQL library. The work on obtaining the results proceeded in two steps. The first step was to work on 1% of the Scopus database data with the snapshot dated August 18, 2021 (from ICSR Lab: 1% of the data volume based on a set of 20,000 publications between 2010 and 2018 and including all publications cited by and citing these publications), using a cluster in standard mode with Databricks Runtime version 11.2, including Apache Spark technology in version 3.3.0, Scala 2.12, and an i3.2xlarge instance with 61 Gbyte memory, eight cores, one to four workers for worker type, and an i3.xlarge instance with 30.5 Gbyte

**Table 1.** The population for 1990–2021: major characteristics

| | | Female scientists | | | Male scientists | | | Total | | |
|---|---|---|---|---|---|---|---|---|---|---|
| | | ***n*** | **row %** | **col %** | ***n*** | **row %** | **col %** | ***N*** | **row %** | **col %** |
| Discipline | Total | 1,645,860 | 38.15 | 100 | 2,668,806 | 61.85 | 100 | 4,314,666 | 100 | 100 |
| | AGRI | 104,805 | 39.98 | 6.37 | 157,318 | 60.02 | 5.89 | 262,123 | 100 | 6.08 |
| | BIO | 328,806 | 46.26 | 19.98 | 381,963 | 53.74 | 14.31 | 710,769 | 100 | 16.47 |
| | CHEM | 87,608 | 30.16 | 5.32 | 202,843 | 69.84 | 7.60 | 290,451 | 100 | 6.73 |
| | CHEMENG | 4,294 | 23.06 | 0.26 | 14,330 | 76.94 | 0.54 | 18,624 | 100 | 0.43 |
| | COMP | 16,191 | 16.59 | 0.98 | 81,414 | 83.41 | 3.05 | 97,605 | 100 | 2.26 |
| | EARTH | 34,042 | 27.62 | 2.07 | 89,221 | 72.38 | 3.34 | 123,263 | 100 | 2.86 |
| | ENER | 3,255 | 19.09 | 0.20 | 13,793 | 80.91 | 0.52 | 17,048 | 100 | 0.40 |
| | ENG | 24,992 | 11.52 | 1.52 | 191,978 | 88.48 | 7.19 | 216,970 | 100 | 5.03 |
| | ENVIR | 35,867 | 38.35 | 2.18 | 57,661 | 61.65 | 2.16 | 93,528 | 100 | 2.17 |
| | IMMU | 26,805 | 53.24 | 1.63 | 23,547 | 46.76 | 0.88 | 50,352 | 100 | 1.17 |
| | MATER | 26,227 | 26.16 | 1.59 | 74,043 | 73.84 | 2.77 | 100,270 | 100 | 2.32 |
| | MATH | 11,915 | 20.15 | 0.72 | 47,206 | 79.85 | 1.77 | 59,121 | 100 | 1.37 |
| | MED | 836,890 | 45.44 | 50.85 | 1,005,040 | 54.56 | 37.66 | 1,841,930 | 100 | 42.69 |
| | NEURO | 40,961 | 47.20 | 2.49 | 45,819 | 52.80 | 1.72 | 86,780 | 100 | 2.01 |
| | PHARM | 15,641 | 41.35 | 0.95 | 22,183 | 58.65 | 0.83 | 37,824 | 100 | 0.88 |
| | PHYS | 47,561 | 15.44 | 2.89 | 260,447 | 84.56 | 9.76 | 308,008 | 100 | 7.14 |
| OECD country (TOP 10) | USA | 540,501 | 39.73 | 32.84 | 819,882 | 60.27 | 30.72 | 1,360,383 | 100 | 31.53 |
| | Japan | 92,601 | 19.28 | 5.63 | 387,599 | 80.72 | 14.52 | 480,200 | 100 | 11.13 |
| | Germany | 118,509 | 33.49 | 7.20 | 235,312 | 66.51 | 8.82 | 353,821 | 100 | 8.20 |
| | UK | 116,285 | 39.49 | 7.07 | 178,187 | 60.51 | 6.68 | 294,472 | 100 | 6.82 |
| | Italy | 119,688 | 50.36 | 7.27 | 117,960 | 49.64 | 4.42 | 237,648 | 100 | 5.51 |
| | France | 93,770 | 42.07 | 5.70 | 129,110 | 57.93 | 4.84 | 222,880 | 100 | 5.17 |
| | Canada | 68,983 | 42.75 | 4.19 | 92,393 | 57.25 | 3.46 | 161,376 | 100 | 3.74 |
| | Spain | 71,656 | 48.13 | 4.35 | 77,233 | 51.87 | 2.89 | 148,889 | 100 | 3.45 |
| | Australia | 50,652 | 44.79 | 3.08 | 62,425 | 55.21 | 2.34 | 113,077 | 100 | 2.62 |
| | South Korea | 19,886 | 19.32 | 1.21 | 83,038 | 80.68 | 3.11 | 102,924 | 100 | 2.39 |

**Table 2.** The subpopulation for 2021: major characteristics

| | | Female scientists | | | Male scientists | | | Total | | |
|---|---|---|---|---|---|---|---|---|---|---|
| | | *n* | row % | col % | *n* | row % | col % | *N* | row % | col % |
| Academic age group | Total | 579,399 | 38.55 | 100 | 923,393 | 61.45 | 100 | 1,502,792 | 100 | 100 |
| | 5 and less | 148,749 | 46.26 | 25.67 | 172,795 | 53.74 | 18.71 | 321,544 | 100 | 21.40 |
| | 6–10 | 149,875 | 43.47 | 25.87 | 194,936 | 56.53 | 21.11 | 344,811 | 100 | 22.94 |
| | 11–15 | 102,419 | 40.52 | 17.68 | 150,366 | 59.48 | 16.28 | 252,785 | 100 | 16.82 |
| | 16–20 | 71,335 | 36.73 | 12.31 | 122,878 | 63.27 | 13.31 | 194,213 | 100 | 12.92 |
| | 21–25 | 45,297 | 32.74 | 7.82 | 93,052 | 67.26 | 10.08 | 138,349 | 100 | 9.21 |
| | 26–30 | 30,302 | 28.86 | 5.23 | 74,698 | 71.14 | 8.09 | 105,000 | 100 | 6.99 |
| | 31–35 | 17,736 | 24.83 | 3.06 | 53,682 | 75.17 | 5.81 | 71,418 | 100 | 4.75 |
| | 36–40 | 8,432 | 20.58 | 1.46 | 32,541 | 79.42 | 3.52 | 40,973 | 100 | 2.73 |
| | 41–45 | 3,833 | 17.27 | 0.66 | 18,357 | 82.73 | 1.99 | 22,190 | 100 | 1.48 |
| | 46–50 | 1,421 | 12.35 | 0.25 | 10,088 | 87.65 | 1.09 | 11,509 | 100 | 0.77 |
| Discipline | AGRI | 42,657 | 40.13 | 7.36 | 63,645 | 59.87 | 6.89 | 106,302 | 100 | 7.07 |
| | BIO | 92,185 | 43.27 | 15.91 | 120,854 | 56.73 | 13.09 | 213,039 | 100 | 14.18 |
| | CHEM | 22,450 | 30.21 | 3.87 | 51,862 | 69.79 | 5.62 | 74,312 | 100 | 4.94 |
| | CHEMENG | 1,287 | 24.98 | 0.22 | 3,865 | 75.02 | 0.42 | 5,152 | 100 | 0.34 |
| | COMP | 6,449 | 18.20 | 1.11 | 28,986 | 81.80 | 3.14 | 35,435 | 100 | 2.36 |
| | EARTH | 14,446 | 27.87 | 2.49 | 37,390 | 72.13 | 4.05 | 51,836 | 100 | 3.45 |
| | ENER | 1,527 | 20.28 | 0.26 | 6,004 | 79.72 | 0.65 | 7,531 | 100 | 0.50 |
| | ENG | 9,029 | 13.82 | 1.56 | 56,326 | 86.18 | 6.10 | 65,355 | 100 | 4.35 |
| | ENVIR | 14,688 | 40.15 | 2.54 | 21,892 | 59.85 | 2.37 | 36,580 | 100 | 2.43 |
| | IMMU | 6,949 | 50.03 | 1.20 | 6,940 | 49.97 | 0.75 | 13,889 | 100 | 0.92 |
| | MATER | 10,257 | 27.09 | 1.77 | 27,601 | 72.91 | 2.99 | 37,858 | 100 | 2.52 |
| | MATH | 4,653 | 20.02 | 0.80 | 18,590 | 79.98 | 2.01 | 23,243 | 100 | 1.55 |
| | MED | 318,792 | 46.14 | 55.02 | 372,166 | 53.86 | 40.30 | 690,958 | 100 | 45.98 |
| | NEURO | 13,873 | 43.76 | 2.39 | 17,833 | 56.24 | 1.93 | 31,706 | 100 | 2.11 |
| | PHARM | 3,190 | 45.98 | 0.55 | 3,748 | 54.02 | 0.41 | 6,938 | 100 | 0.46 |
| | PHYS | 16,967 | 16.53 | 2.93 | 85,691 | 83.47 | 9.28 | 102,658 | 100 | 6.83 |
| OECD country (TOP 10) | USA | 176,646 | 40.63 | 30.49 | 258,155 | 59.37 | 27.96 | 434,801 | 100 | 28.93 |
| | Japan | 22,331 | 18.15 | 3.85 | 100,695 | 81.85 | 10.90 | 123,026 | 100 | 8.19 |
| | Germany | 36,659 | 32.19 | 6.33 | 77,212 | 67.81 | 8.36 | 113,871 | 100 | 7.58 |

**Table 2.** (*continued*)

| | Female scientists | | | Male scientists | | | Total | | |
|---|---|---|---|---|---|---|---|---|---|
| | *n* | row % | col % | *n* | row % | col % | *N* | row % | col % |
| Italy | 51,171 | 49.21 | 8.83 | 52,821 | 50.79 | 5.72 | 103,992 | 100 | 6.92 |
| UK | 40,328 | 38.88 | 6.96 | 63,392 | 61.12 | 6.87 | 103,720 | 100 | 6.90 |
| France | 31,657 | 39.74 | 5.46 | 47,996 | 60.26 | 5.20 | 79,653 | 100 | 5.30 |
| Spain | 29,067 | 46.89 | 5.02 | 32,925 | 53.11 | 3.57 | 61,992 | 100 | 4.13 |
| Canada | 24,022 | 42.36 | 4.15 | 32,685 | 57.64 | 3.54 | 56,707 | 100 | 3.77 |
| Australia | 21,160 | 44.49 | 3.65 | 26,396 | 55.51 | 2.86 | 47,556 | 100 | 3.16 |
| South Korea | 7,903 | 19.31 | 1.36 | 33,034 | 80.69 | 3.58 | 40,937 | 100 | 2.72 |

memory and four cores for the driver type. Test runs of the scripts covered 1% of the data, with the goal of optimizing the time and cost of the performed calculations.

After reviewing the correctness of the scripts, the final run was performed. The operation was carried out on a 100% Scopus database with a snapshot date October 21, 2022, using a cluster in standard mode with Databricks Runtime version 11.2 ML with Apache Spark technology version 3.3.0, Scala 2.12, and an instance i3.2xlarge with 61 Gbyte memory, eight cores, one to six workers for worker type, and an instance c4.2xlarge with 15 Gbyte memory and four cores for the driver type. The execution time for the entire script took 1.13 hours; this operation was launched on November 22, 2022.

#### 2.2.1. Gender determination

To obtain the gender of the scientists in the population, the gender data established by the ICSR Lab platform were first used ($N_{author}$ = 34,596,581). Then, only scientists who had a defined gender (man/woman) with a gender probability score greater than or equal to 0.85 were included ($N_{author}$ = 21,508,029). To assign gender to an author, the ICSR Lab used Elsevier's solution, which used the Namsor tool. Determining gender was based on three characteristics: author's first name, author's last name, and author's first country. The author's first country was determined based on the author's dominant country in their first publication year, which was based on output in the Scopus database. For authors who had more than one dominant country, the observation was not assigned a value. The Namsor tool returned gender and gender probability score (Elsevier, 2020, pp. 122–123).

#### 2.2.2. Discipline determination

To obtain the dominant discipline of scientists in the population, a set of publications from the Scopus database was used ($N_{pub}$ = 85,585,123; $N_{author}$ = 43,632,099). Publications were from 2021 and before and were restricted by source and type of publication: (a) journal article and (b) conference paper in a book or journal ($N_{pub}$ = 60,987,987; $N_{author}$ = 36,379,221). From the table of publications, the columns with publications' identifiers, authors' identifiers, and cited references were selected. Each cited reference ($N_{citedreference}$ = 1,434,621,669) was accompanied by its discipline, as assigned by the discipline of the journal in which it appeared. The disciplines assigned to a cited reference were based on the four-digit ASJC code used by the Scopus database. To switch to a two-digit classification, unique disciplines were selected, here based on the first two digits of the four-digit value. Then, for each author, the number of cited references was counted for all disciplines referenced by the author, excluding the "multidisciplinary" discipline. For each author, the discipline with the highest number of cited references (modal value) was selected. A table containing the author's identifier and their dominant discipline was obtained. For the described summary, there could have been cases in which an author had several dominant disciplines or no disciplines (included $N_{author}$ = 26,706,031). Here, authors who had more than one dominant discipline or no discipline were removed from the table (removed $N_{author}$ = 9,673,190). Authors were removed, among other reasons, because the cited references from their articles may have referred to journals outside the Scopus database or because there was an equal number of cited references to different disciplines. Subsequently, the table was restricted to only authors with an assigned discipline from the STEMM group, resulting in the final number ($N_{author}$ = 24,425,447).

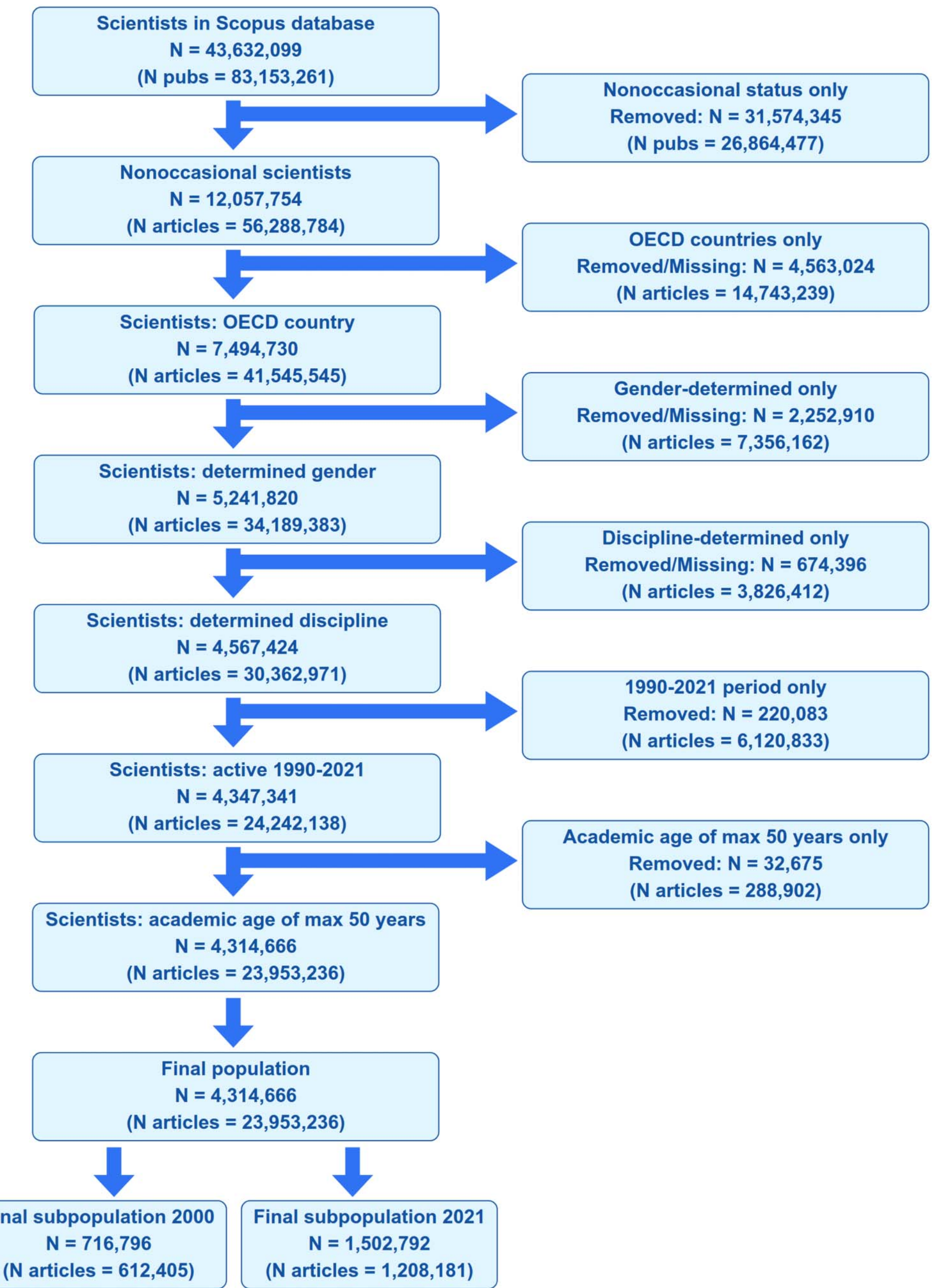

**Figure 1.** Flowchart: stages in constructing the population and two subpopulations.

#### 2.2.3. Determining the country of affiliation

Publications were from 2021 or earlier and were restricted by source and type: (a) journal article and (b) conference paper in a book or journal. From the table of publications, columns with publications' identifiers, authors' identifiers, and countries for each author of the publication were selected. Then, for each author, the number of countries that the scientist indicated in all their publications was counted. For each author, the country with the highest number of references (modal value) was selected. For the described summary, there may have been cases in which an author had several countries (included $N_{author}$ = 31,332,750). For this purpose, authors who had more than one country or no countries were removed from the table (removed $N_{author}$ = 5,046,471). The table was then filtered to include scientists from 38 OECD countries. The final number was ($N_{author}$ = 19,296,388).

#### 2.2.4. Determining scientists' nonoccasional status

Under the proposed definition, a nonoccasional scientist has at least three research articles (as defined above) in their output. The publications were from 2021 or before and were limited by the same source and type of publication as above. Columns containing publications' identifiers and authors' identifiers were selected from the table of publications. For each author, the number of publications was counted. The table was then filtered to include scientists who had a minimum of three publications ($N_{author}$ = 12,057,755).

#### 2.2.5. Determining academic age

Finally, to obtain the academic age of the scientists in the population, the same set of publications from the Scopus database was used, and the publications were from 2021 or before. Author identifiers and year of publication were selected from the table. For each author, the year of the first and last publication (of any type) was determined. Then, the number of years of authors' research activities (distance from the first to last publication in years) was calculated according to the following formula: year of the last publication – year of the first publication + 1. Authors who had more than 50 years of research activities were removed from the table (included $N_{author}$ = 43,568,252; removed $N_{author}$ = 63,847). Then, for the authors included in the study ($N_{author}$ = 4,314,666; i.e., the final population) that contained the years of academic activity defined for publications, the academic age in a given publication year was determined according to the following formula: publication year – year of first publication + 1. Based on the value of academic age, an author was assigned to an age group according to 10 ranges: 5 and less, 6–10, 11–15, 16–20, 21–25, 26–30, 31–35, 36–40, 41–45, and 46–50.

#### 2.2.6. List of STEMM disciplines

We focused on all 16 STEMM disciplines, as defined by the journal classification system used in the Scopus database (All Science Journal Classification, ASJC): AGRI, agricultural and biological sciences; BIO, biochemistry, genetics, and molecular biology; CHEMENG, chemical engineering; CHEM, chemistry; COMP, computer science; EARTH, earth and planetary sciences; ENER, energy; ENG, engineering; ENVIR, environmental science; IMMU, immunology and microbiology; MATER, materials science; MATH, mathematics; MED medicine, NEURO, neuroscience; PHARM, pharmacology, toxicology, and pharmaceutics; and PHYS, physics and astronomy.

## 3. RESULTS

To study the gender distribution of the scientific workforce by age group, we used two complementary approaches we termed "horizontal" and "vertical."

1. A horizontal approach: analyzing the gender distribution of scientists horizontally within the same age groups. For each discipline, for each of the 10 5-year age groups, the percentages of male and female scientists totaled 100%.
2. A vertical approach: analyzing the gender distribution of scientists vertically—separately male and separately female scientists—across all age groups. For each discipline, there was 100% of male and 100% of female scientists who were differently distributed across the 10 age groups.

Parts of our study have been based on a longitudinal research design in a broader sense, which requires a short methodological explanation. In longitudinal studies in a narrow sense, data are collected at multiple points in time from the same group of participants; we used this narrow approach in a recent study of 2,326 Polish full professors, tracing their promotions, publications, and productivity classes over a period of 40 years (Kwiek & Roszka, 2023a). In classical definitions, longitudinal research concerns data collection and analysis over time, and it is a broad term that describes a family of methods: Specifically, longitudinal research includes repeated cross-sectional studies, prospective studies, and retrospective studies (Menard, 2002, pp. 2–3). As a minimum, any longitudinal design would permit the measurement of differences or a change in a variable from one period to another. In this broader sense, longitudinal research is research in which data are collected for each item or variable for two or more distinct time periods; the subjects or cases analyzed are the same (or at least comparable) from one period to the next; and the analysis involves some comparison of data between or among periods (Menard, 2002, p. 2).

Our study represents both a cross-sectional design (in its analyses of a single point in time, 2021) and longitudinal design in a broader sense, in its repeated cross-sectional design variation (analyzing two points in time, 2000 and 2021, and the trend in 1990–2021, following the idea that cross-sectional data are repeated over time with a high level of consistency between questions; Ruspini, 1999). Our sets of cases, scientists with their microdata, for each period are not entirely different: to some extent, they overlap (for scientists active for a longer period of time). Our microdata are at the individual level of scientists, which means that their individual-level records contain the same variables measured at several different time points. For this reason, they were pooled to form a single data file: This increased the sample size and also introduced a temporal dimension, as suggested in the literature (Ruspini, 1999, p. 222).

This section is divided into the subsections on general results (Section 3.1), results from horizontal (Section 3.2) and vertical (Section 3.3) perspectives, which include both cross-sectional and longitudinal data, and the results of the trend analysis for the 1990–2021 period based on a longitudinal data set (Section 3.4).

### 3.1. General Results

Although the analysis of the changing numbers of male and female scientists over time may be distorted by the inability to distinguish between an expansion in numbers of scientists and in numbers of journals indexed in large bibliometric data sets, in contrast, the changing relative presence of female scientists is traceable. Although the increasing number of publishing scientists over time correlated with the increasing coverage in Scopus, the percentages of publishing male and female scientists were independent of journal coverage. Consequently, although the *number* of publishing scientists changing over time was not a reliable measure of the changing women's participation in global science, the *percentages* of male scientists and female scientists adequately reflected the changes in the global academic workforce.

In 2021, 45.98% of the global scientific workforce in STEMM (as defined in this research, especially in the 38 OECD countries and with a nonoccasional publishing status in Scopus) was involved in medical research, with 690,958 scientists in medicine, followed by biochemistry, genetics, and molecular biology (213,039). In 2021, there were 1.5 million scientists, as defined in our population, with 923,000 men and 579,000 women (38.55%). The majority of female scientists (63.09%) were concentrated in six countries: the United States, Italy, the United Kingdom, Germany, France, and Spain. Over 70% of female scientists were in medicine and biochemistry, genetics, and molecular biology. Immunology and microbiology had the highest share of female scientists (50.03%), followed by several fields with over 40% female representation (e.g., AGRI, ENVIR, BIO, NEURO, PHARM, and MED). In contrast, engineering, physics and astronomy, computer science, and mathematics had 20% (or less) female representation (see Figure 2).

### 3.2. Results: A Horizontal Approach

#### *3.2.1. A cross-sectional view (2021): All age groups horizontally*

Disciplines at a single point in time (2021) were populated by the scientists of different age groups and genders. Figure 3 shows the percentage of female scientists across disciplines by age group. We generally observed the results of a huge inflow of female scientists (who were present in 2021) to most disciplines in the past years and decades: For younger generations working in 2021, the percentages of female scientists were substantially higher than for older generations.

**Table 3.** The subpopulation for 2021 by discipline and gender, as sorted by the number of male scientists (in descending order)

| Discipline | Total | Female scientists | Male scientists | Percentage female | Percentage male |
|---|---|---|---|---|---|
| MED | 690,958 | 318,792 | 372,166 | 46.14 | 53.86 |
| BIO | 213,039 | 92,185 | 120,854 | 43.27 | 56.73 |
| AGRI | 106,302 | 42,657 | 63,645 | 40.13 | 59.87 |
| PHYS | 102,658 | 16,967 | 85,691 | 16.53 | 83.47 |
| CHEM | 74,312 | 22,450 | 51,862 | 30.21 | 69.79 |
| ENG | 65,355 | 9,029 | 56,326 | 13.82 | 86.18 |
| EARTH | 51,836 | 14,446 | 37,390 | 27.87 | 72.13 |
| MATER | 37,858 | 10,257 | 27,601 | 27.09 | 72.91 |
| ENVIR | 36,580 | 14,688 | 21,892 | 40.15 | 59.85 |
| COMP | 35,435 | 6,449 | 28,986 | 18.20 | 81.80 |
| NEURO | 31,706 | 13,873 | 17,833 | 43.76 | 56.24 |
| MATH | 23,243 | 4,653 | 18,590 | 20.02 | 79.98 |
| IMMU | 13,889 | 6,949 | 6,940 | 50.03 | 49.97 |
| ENER | 7,531 | 1,527 | 6,004 | 20.28 | 79.72 |
| PHARM | 6,938 | 3,190 | 3,748 | 45.98 | 54.02 |
| CHEMENG | 5,152 | 1,287 | 3,865 | 24.98 | 75.02 |
| **Total** | **1,502,792** | **579,399** | **923,393** | **38.55** | **61.45** |

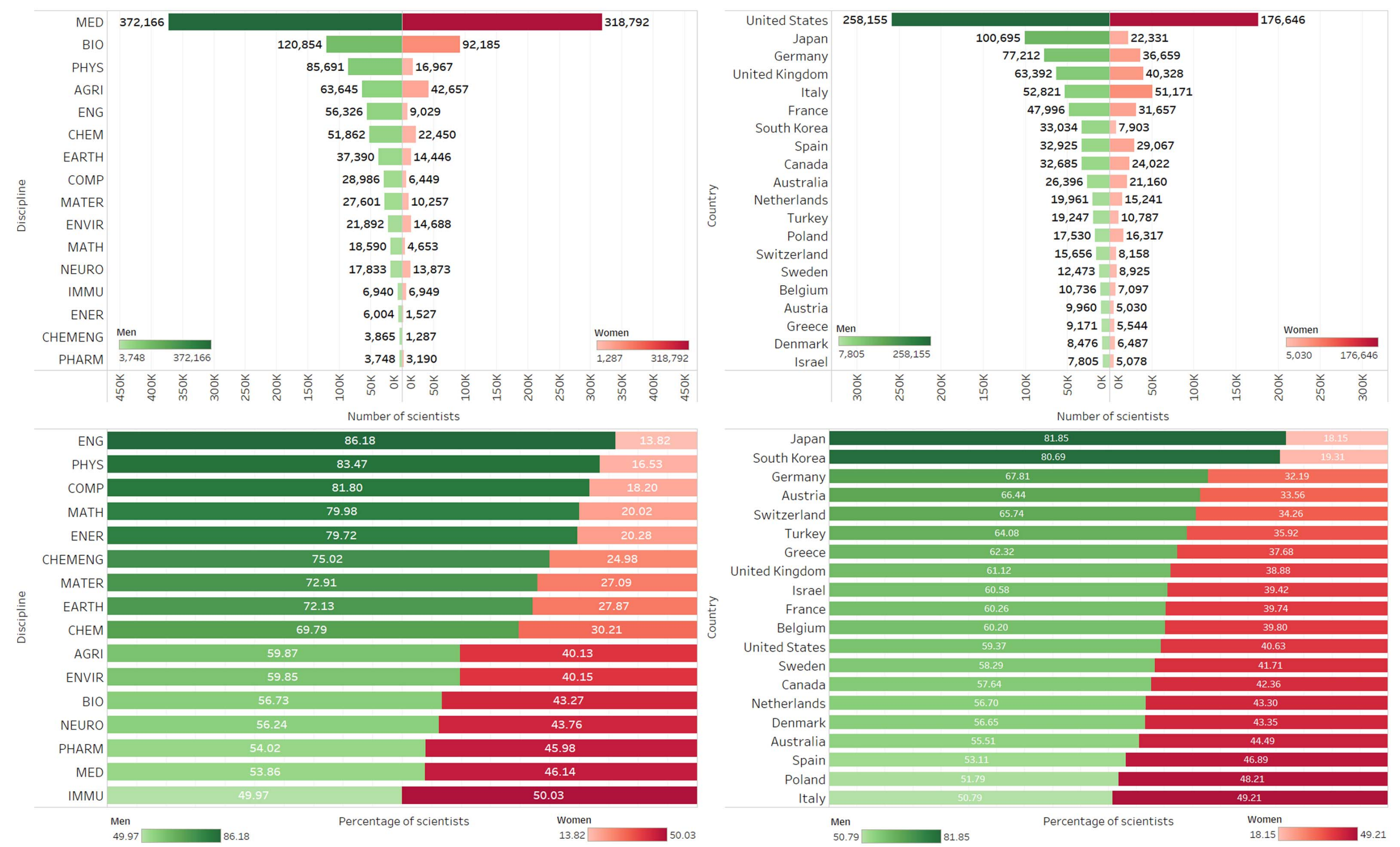

**Figure 2.** The number of publishing nonoccasional STEMM scientists in 38 OECD countries by discipline and gender (left top) and by country (20 biggest systems only) and gender (right top). The share by discipline and gender (left bottom) and by country (20 biggest OECD systems only) and gender (right bottom) (in %), 2021 ($N$ = 1,502,792).

Generally expecting ever more female scientists across all STEMM disciplines moving down the age groups, we assessed ongoing changes based on a snapshot (2021), especially examining the youngest age groups. MED and BIO showed a structure in which, for every successive lower age group in 2021, a higher share of female scientists was observed. PHYS, COMP, ENG, and MATH, termed the *Big Four* in this paper, which have been traditionally male-dominated disciplines comprising about 262,000 scientists in our population (15.09%; including merely 30,649 women), in contrast, showed a stable structure in which, for every successive lower age group in 2021, a similar (or only slightly higher) share of female scientists was observed. These two contrasting demographic patterns showed different inflows of young female scientists to disciplines in the past: huge and increasing versus small and stable. This can be compared with MATH and BIO: in a single year of interest, with the most recent data available, the share of very young, young, and middle-aged women is almost the same; in contrast, the share of women in the same age groups for BIO increases continually with every age group.

The current global disciplinary distribution of young women in science is consequential for gender parity in science in the future, despite the high attrition rate among young scientists generally and young female scientists in particular (1 in 10; see Boothby, Milojevic et al., 2022). The current young cohorts will be middle-aged cohorts within a decade, and the current oldest cohorts will disappear from the publishing enterprise and exit from academic work, with new challenges for disciplines continuously heavily male dominated.

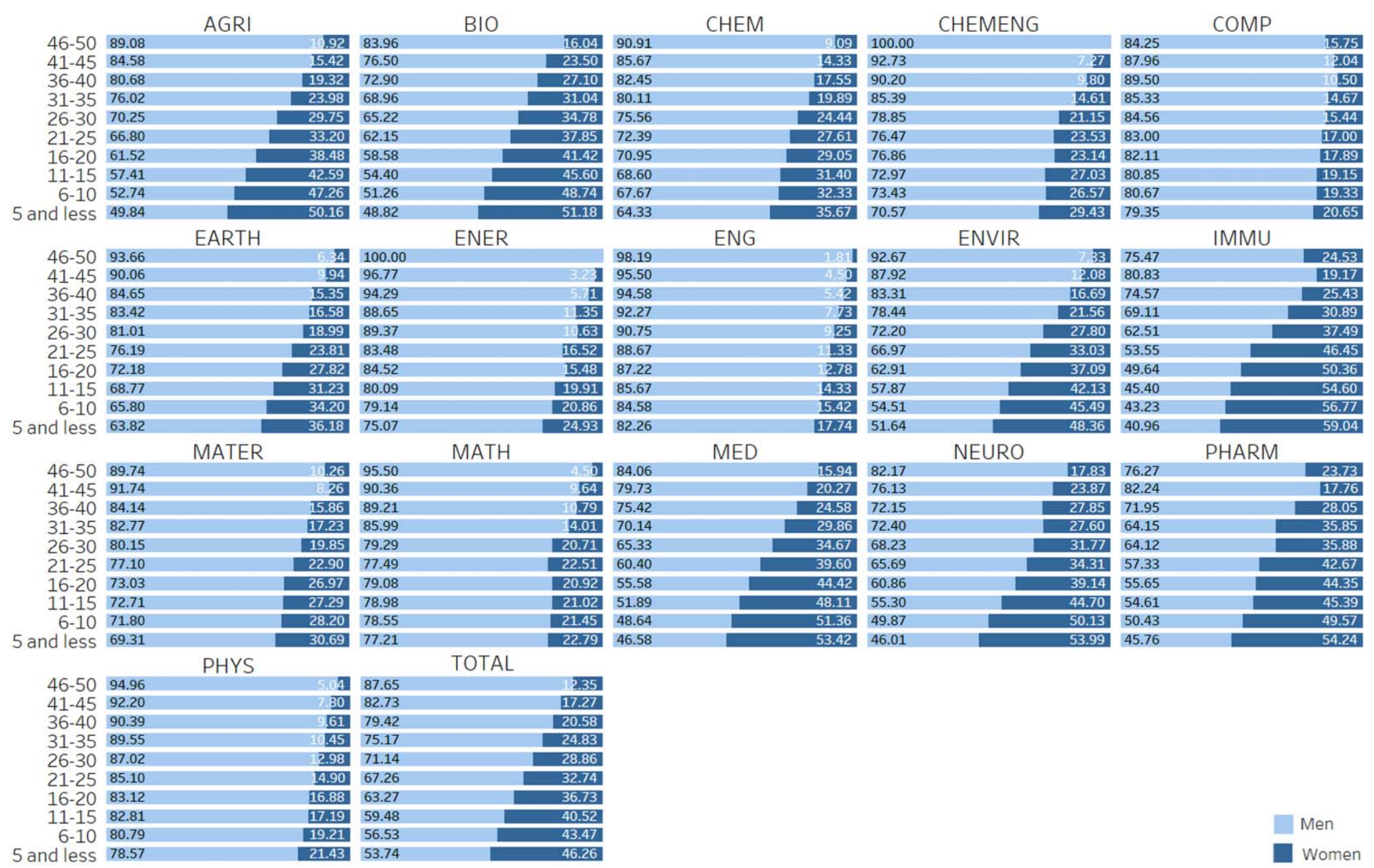

**Figure 3.** Ever-increasing participation of women in younger generations of scientists, with a few exceptions (e.g., COMP, MATH). Horizontal approach: distribution of publishing nonoccasional STEMM scientists by discipline, age group, and gender (row percentages: 100% horizontally), 2021 ($N$ = 1,502,792).

Traditional gender-aggregated and age-aggregated data about scientists in general across disciplines, countries, and institutions hide a much more nuanced picture of the changing gender dynamics within and across disciplines and age groups. In this research, we examined the subpopulation of "young" scientists (academic age 10 and years and below: Figure 4).

#### 3.2.2. A comparative horizontal view (2000 vs. 2021)

When comparing women's participation in STEMM disciplines from another perspective of two snapshots of 2000 and 2021 (Figure 5), for all disciplines, the share of female scientists increased, albeit to different degrees. The white lines show the shares of female scientists for the year 2000, and the dark blue bars on the right show this for 2021. For the youngest age group, for all disciplines combined, the share of female scientists increased from one-third to a half (from 34.93% to 50.16%), indicating that the share of male scientists decreased from two-thirds to a half (from 65.07% to 49.84%). Comparing the old age category of 31–35, the share of female scientists increased three times, from 8.12% to 23.98%. From the perspective of two decades, the changes are noticeable across all disciplines—although in most cases, they can be described as small scale.

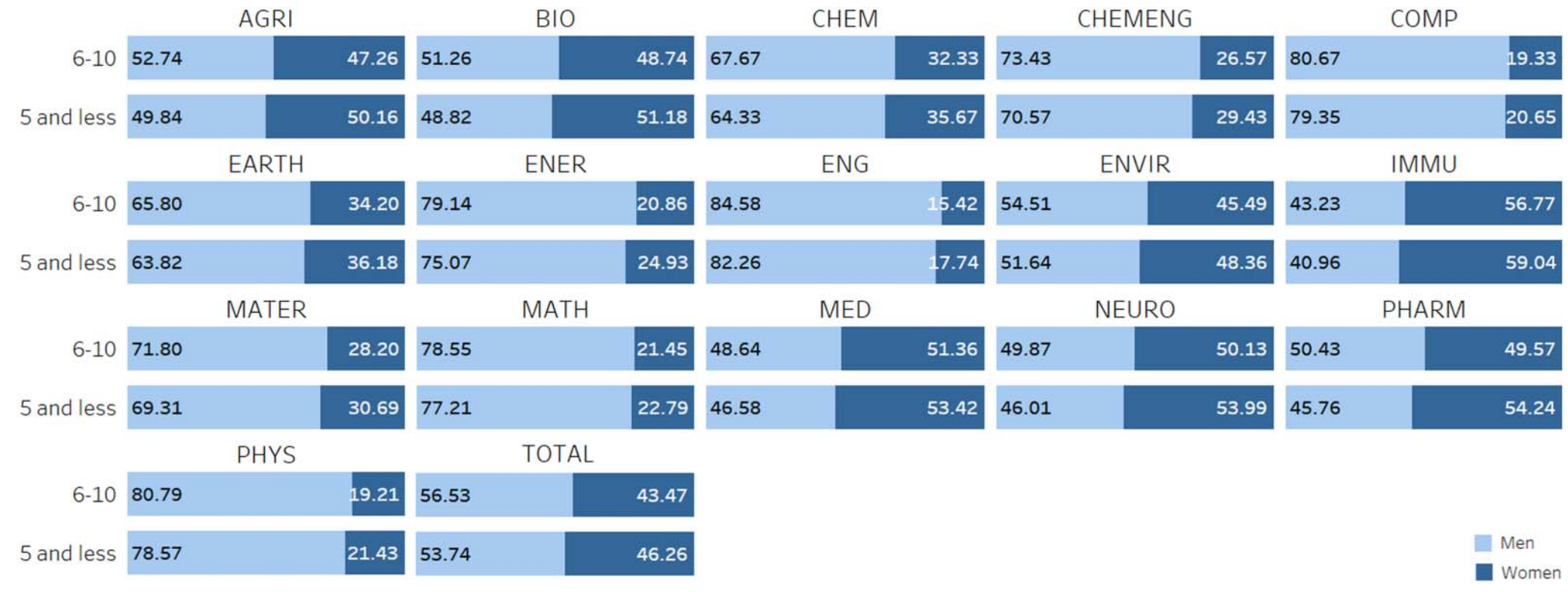

**Figure 4.** Zooming in on young scientists only. More young men than young women in all but six STEMM disciplines (e.g., MED). Horizontal approach: young scientists only (academic age 10 years and less). Distribution of young publishing nonoccasional STEMM scientists by discipline, age group, and gender (row percentages: 100% horizontally), 2021 ($N$ = 666,355).

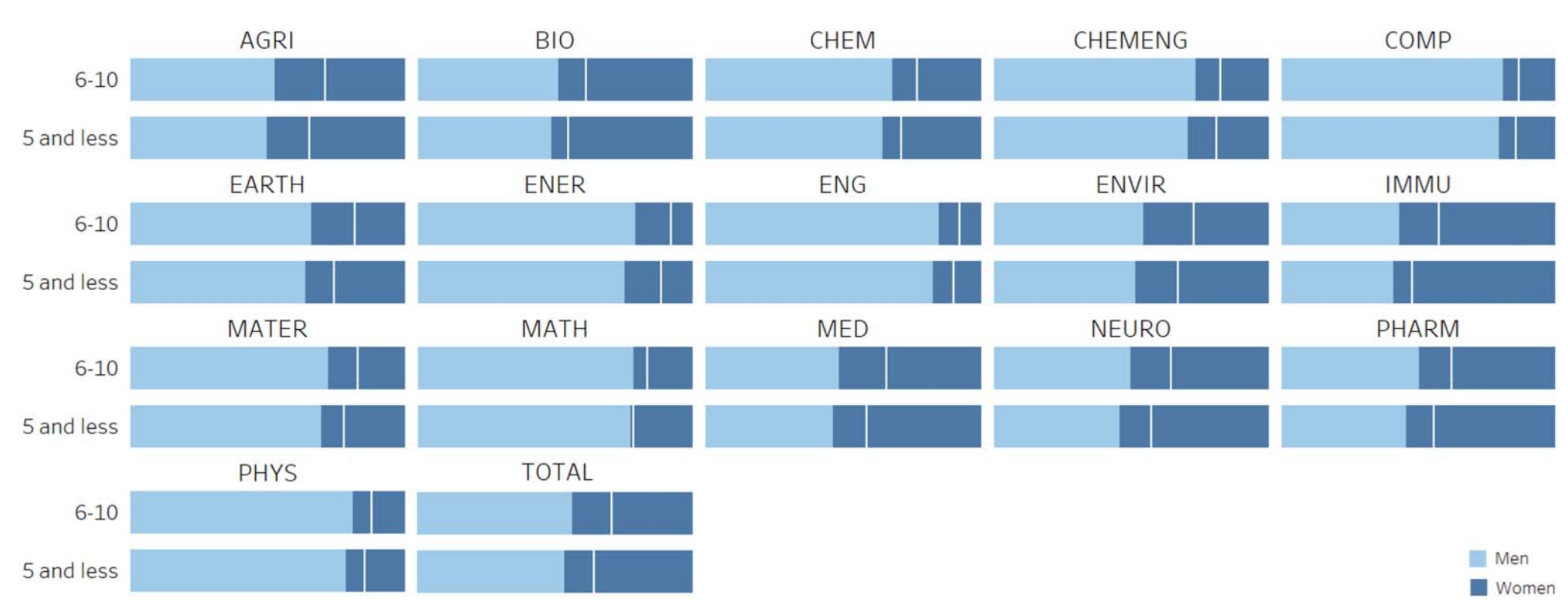

**Figure 5.** The increasing participation of young female scientists for all disciplines over time. Overview of percentage change directions, 2000 vs. 2021: horizontal approach. Zooming in on young scientists only (academic age 10 years or less). Distribution of young publishing nonoccasional STEMM scientists by discipline, age group, and gender; dark blue percentage female scientists 2021, white lines percentage female scientists 2000 (row percentages: 100% horizontally) ($N_{2021}$ = 666,355, $N_{2000}$ = 437,113).

#### 3.2.3. *The decreasing isolation of female scientists in the Big Four math-intensive disciplines*

We compared the share of young and old female and male scientists across disciplines to show gender differences. Among the young scientists, the share of female scientists in several disciplines was about half (e.g., BIO and MED), and among old scientists, the share was much lower (Table 4). In some disciplines, the share of old female scientists was about 10% or lower, with gender differences being at least 10-fold (e.g., ENG and PHYS: 6.31% and 9.21%, respectively).

In many institutions, old female scientists were not merely minorities: They were tokens (or single, exemplary scientists representing all female scientists; see Kanter (1977); on the role of microlevel departmental climates, see Fox and Nikivincze (2021)). However, the isolation of young female scientists in COMP, ENG, MATH, and PHYS decreased at least twice, with higher visibility for young cohorts. Younger age groups had more female scientists and higher percentages across all disciplines, including male-dominated ones (ENG, MATH, PHYS) and those closer to gender parity (MED, AGRI, BIO). Female scientists were more present in numbers and percentages in younger age groups. For all disciplines except six (AGRI, BIO, IMMU, MED, NEURO, and PHARM), there were more of the youngest male scientists than youngest female scientists, and many more old male scientists than old female scientists. Table 4 shows a general increase in the percentage of female scientists among the younger age groups (10 years or less of experience) compared with the older age groups (31–50 years of experience) across all disciplines. This suggests a growing trend in women's participation in science.

However, in the context of scientists from different age groups working at the same time (2021) in the Big Four, the isolation of young female scientists has decreased significantly compared with the isolation of old female scientists. In 2021, for these four disciplines, the percentage of female scientists in the younger generations was at least twice that in older generations: for instance, in engineering, young female scientists made up 16.47% of the total, compared with only 6.31% for the older cohorts (Table 5).

**Table 4.** The frequencies and percentages of female scientists among publishing nonoccasional STEMM scientists by discipline in the two age cohorts (the young and the old), 2021

| Young scientists (10 years of publishing experience or less) | | | | Old scientists (31–50 years of publishing experience) | | | |
|---|---|---|---|---|---|---|---|
| **Discipline** | **All young scientists** | **Young female scientists** | **% female scientists** | **Discipline** | **All old scientists (31–50)** | **Old female scientists** | **% female scientists** |
| AGRI | 41,954 | 20,389 | 48.60 | AGRI | 10,799 | 2,206 | 20.43 |
| BIO | 89,295 | 44,533 | 49.87 | BIO | 23,377 | 6,422 | 27.47 |
| CHEM | 36,368 | 12,394 | 34.08 | CHEM | 7,582 | 1,313 | 17.32 |
| CHEMENG | 2,523 | 707 | 28.02 | CHEMENG | 455 | 51 | 11.21 |
| COMP | 12,678 | 2,518 | 19.86 | COMP | 2,642 | 353 | 13.36 |
| EARTH | 18,168 | 6,363 | 35.02 | EARTH | 7,205 | 1,026 | 14.24 |
| ENER | 4,420 | 1,013 | 22.92 | ENER | 252 | 21 | 8.33 |
| ENG | 28,808 | 4,745 | 16.47 | ENG | 4,864 | 307 | 6.31 |
| ENVIR | 16,557 | 7,758 | 46.86 | ENVIR | 2,545 | 458 | 18.00 |
| IMMU | 5,651 | 3,270 | 57.87 | IMMU | 1,587 | 430 | 27.10 |
| MATER | 20,664 | 6,103 | 29.53 | MATER | 2,097 | 323 | 15.40 |
| MATH | 8,327 | 1,835 | 22.04 | MATH | 3,481 | 386 | 11.09 |
| MED | 324,524 | 170,004 | 52.39 | MED | 60,685 | 15,775 | 25.99 |
| NEURO | 14,260 | 7,400 | 51.89 | NEURO | 2,903 | 758 | 26.11 |
| PHARM | 3,341 | 1,741 | 52.11 | PHARM | 744 | 223 | 29.97 |
| PHYS | 38,817 | 7,851 | 20.23 | PHYS | 14,872 | 1,370 | 9.21 |
| **Total** | **666,355** | 298,624 | 44.81 | Total | 146,090 | 31,422 | 21.51 |

This trend of higher female representation in younger cohorts is stronger across disciplines closer to gender parity. For example, in medicine in 2021, young female scientists made up 52.39% of the total (compared with 25.99% for old scientists; and in biochemistry, 49.87% and 27.47%, respectively).

With our individual-level microdata, we can explore further what the isolation of female scientists in STEMM disciplines means in practice. As shown in Table 5, female scientists are more present in numbers and percentages when moving from older to younger generations across the 10 age groups in the same year 2021. This indicates a positive trend toward decreasing isolation of female scientists with every next younger age group. Detailed examples from specific age groups can further emphasize the contrast between the presence of female scientists in younger and older generations in 2021.

In engineering, for instance, in the 36–40 years age group, there were only 84 (nonoccasional, publishing, etc.) female scientists compared with 1,486 male scientists. This shows a stark contrast in representation, with males outnumbering females by more than 17 times. However, in the younger (5 years and less) age group, the gap has narrowed considerably, with 2,316 female engineers and 10,739 male engineers. In this case, the number of male scientists is only about 4.6 times higher than that of female scientists. This example illustrates that the isolation of female scientists in engineering has decreased significantly in younger generations. In physics and astronomy (PHYS), in the 46–50 years age group, there were only 79 female scientists compared with 1,489 male scientists (nearly 19 times difference in this age group). In the younger (5 years and less) age group, however, there were 3,817 female physicists and 13,998 male physicists (only about 3.7 times difference). The academic worlds of young scientists in the Big Four today and 20–30 years ago are amazingly different, with the old today being the young decades ago and surviving in heavily male-dominated environments.

**Table 5.** Zooming in on numbers of the young vs. the old: gender- and age-disaggregated data, distribution of nonoccasional publishing STEMM scientists by selected academic age groups and gender, 2021

| Discipline | Gender | 5 years and less | 6–10 years | 31–35 years | 36–40 years |
|---|---|---|---|---|---|
| AGRI | Female | 9,714 | 10,675 | 1,238 | 647 |
| | Male | 9,652 | 11,913 | 3,925 | 2,702 |
| BIO | Female | 21,139 | 23,394 | 3,463 | 1,757 |
| | Male | 20,161 | 24,601 | 7,692 | 4,726 |
| CHEM | Female | 6,793 | 5,601 | 693 | 380 |
| | Male | 12,253 | 11,721 | 2,792 | 1,785 |
| CHEMENG | Female | 377 | 330 | 32 | 15 |
| | Male | 904 | 912 | 187 | 138 |
| COMP | Female | 1,049 | 1,469 | 231 | 76 |
| | Male | 4,030 | 6,130 | 1,344 | 648 |
| EARTH | Female | 2,732 | 3,631 | 534 | 335 |
| | Male | 4,820 | 6,985 | 2,686 | 1,848 |
| ENER | Female | 557 | 456 | 16 | 4 |
| | Male | 1,677 | 1,730 | 125 | 66 |
| ENG | Female | 2,316 | 2,429 | 198 | 84 |
| | Male | 10,739 | 13,324 | 2,362 | 1,466 |
| ENVIR | Female | 3,807 | 3,951 | 277 | 130 |
| | Male | 4,065 | 4,734 | 1,008 | 649 |
| IMMU | Female | 1,617 | 1,653 | 249 | 104 |
| | Male | 1,122 | 1,259 | 557 | 305 |
| MATER | Female | 3,397 | 2,706 | 193 | 98 |
| | Male | 7,670 | 6,891 | 927 | 520 |
| MATH | Female | 829 | 1,006 | 193 | 112 |
| | Male | 2,808 | 3,684 | 1,185 | 926 |
| MED | Female | 86,100 | 83,904 | 9,217 | 4,005 |
| | Male | 75,065 | 79,455 | 21,655 | 12,289 |
| NEURO | Female | 3,520 | 3,880 | 369 | 227 |
| | Male | 3,000 | 3,860 | 968 | 588 |
| PHARM | Female | 985 | 756 | 128 | 62 |
| | Male | 831 | 769 | 229 | 159 |

**Table 5.** (*continued*)

| Discipline | Gender | 5 years and less | 6–10 years | 31–35 years | 36–40 years |
|---|---|---|---|---|---|
| PHYS | Female | 3,817 | 4,034 | 705 | 396 |
| | Male | 13,998 | 16,968 | 6,040 | 3,726 |
| **Total** | Female | 148,749 | 149,875 | 17,736 | 8,432 |
| | Male | 172,795 | 194,936 | 53,682 | 32,541 |

### 3.3. Results: A Vertical Approach

#### 3.3.1. *A cross-sectional view (2021): All age groups vertically*

Examining the gender composition within disciplines, we found that, in the majority of disciplines (nine), most female scientists were in the two young age groups; that is, with no more than 10 years of academic experience (Figure 6). Young female scientists dominated (> 50%) among all female scientists in disciplines like CHEM, ENG, or MED. Thus, the inflow of (publishing nonoccasional) female scientists in the past decade or so in these disciplines has been massive. The lowest share of young female scientists among all female scientists—or the weakest inflow (< 40%)—was for COMP and MATH. In all disciplines combined (Total), the share of young female scientists among all female scientists reached 51.54%, and the share of young male scientists among all male scientists was considerably lower and reached 39.82%. The emergent picture supports narratives of an increasing number of young women in science: of all the women currently present in global science, more than half had no more than 10 years of publishing experience (see the details in Figure 7).

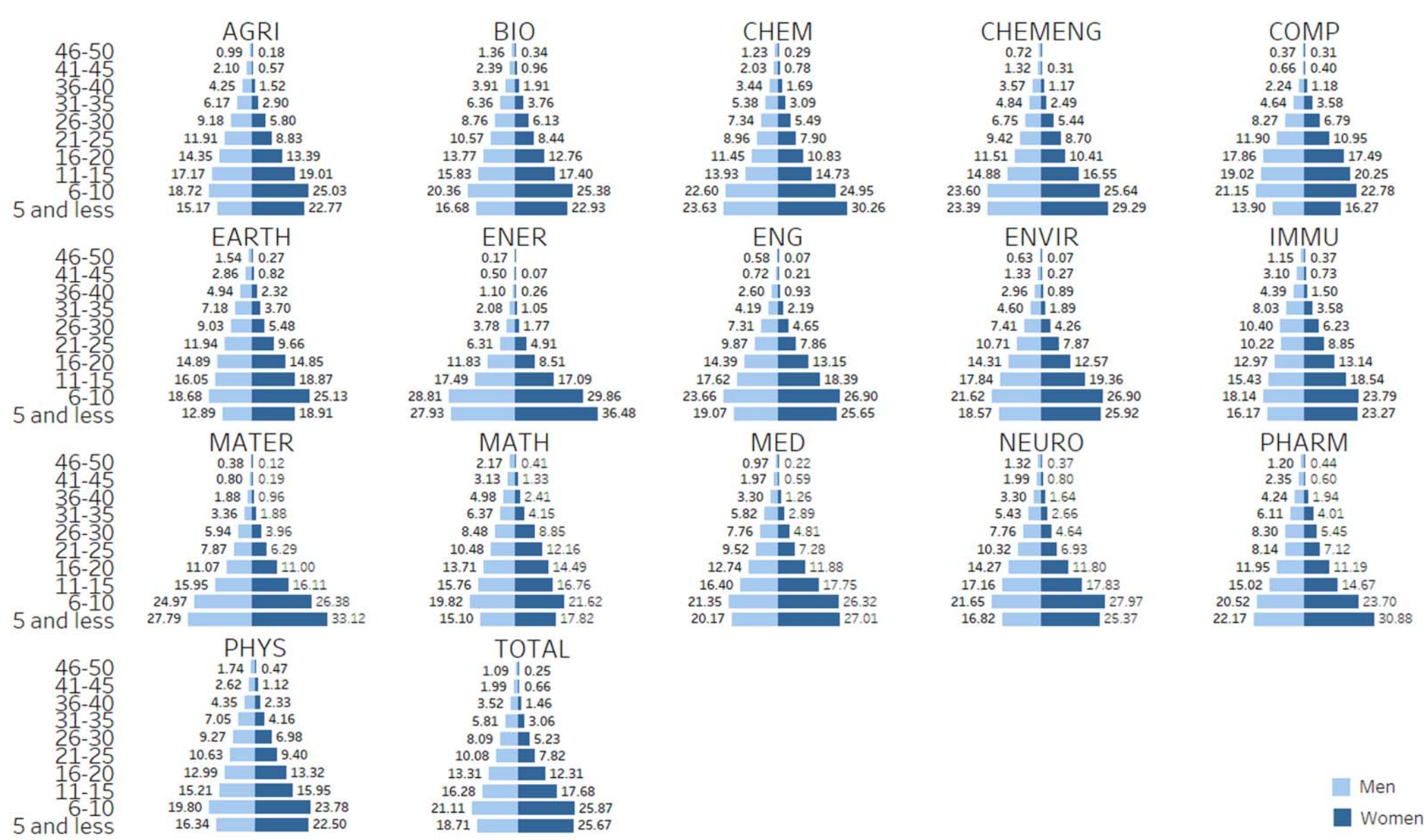

**Figure 6.** Young women in STEMM: In most disciplines, the majority of women belong to the two youngest age groups. Vertical approach: distribution of publishing nonoccasional STEMM scientists by discipline, age group, and gender (column percentages: 100% vertically, for all age groups combined), 2021 ($N$ = 1,502,792).

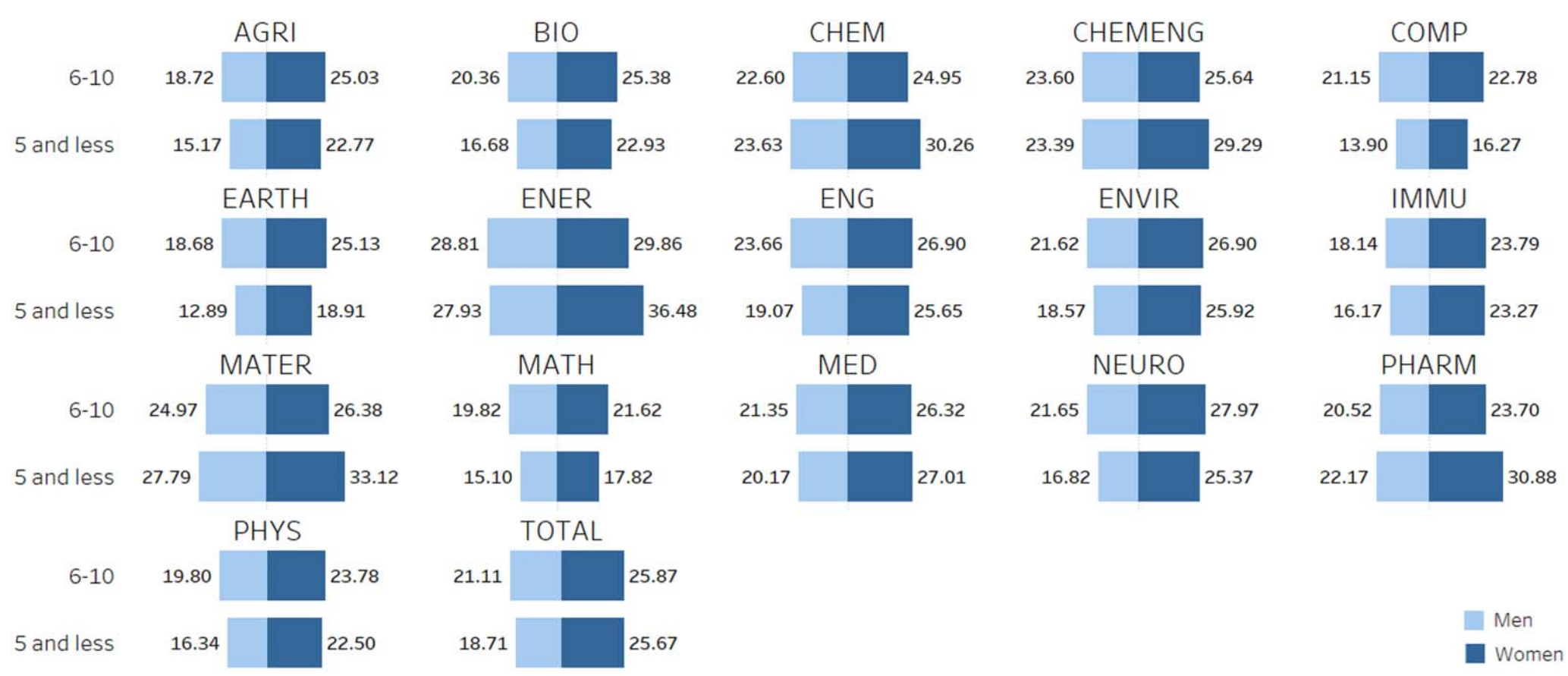

**Figure 7.** Zooming in on young scientists only. Higher concentration of young women than young men across all disciplines. Vertical approach: zooming in on young scientists only (academic age 10 and less)—distribution of publishing nonoccasional STEMM scientists by discipline, age group, and gender (column percentages, vertically: percentage of young female scientists among all women, and young men among all men; women in dark blue), 2021 ($N$ = 666,355).

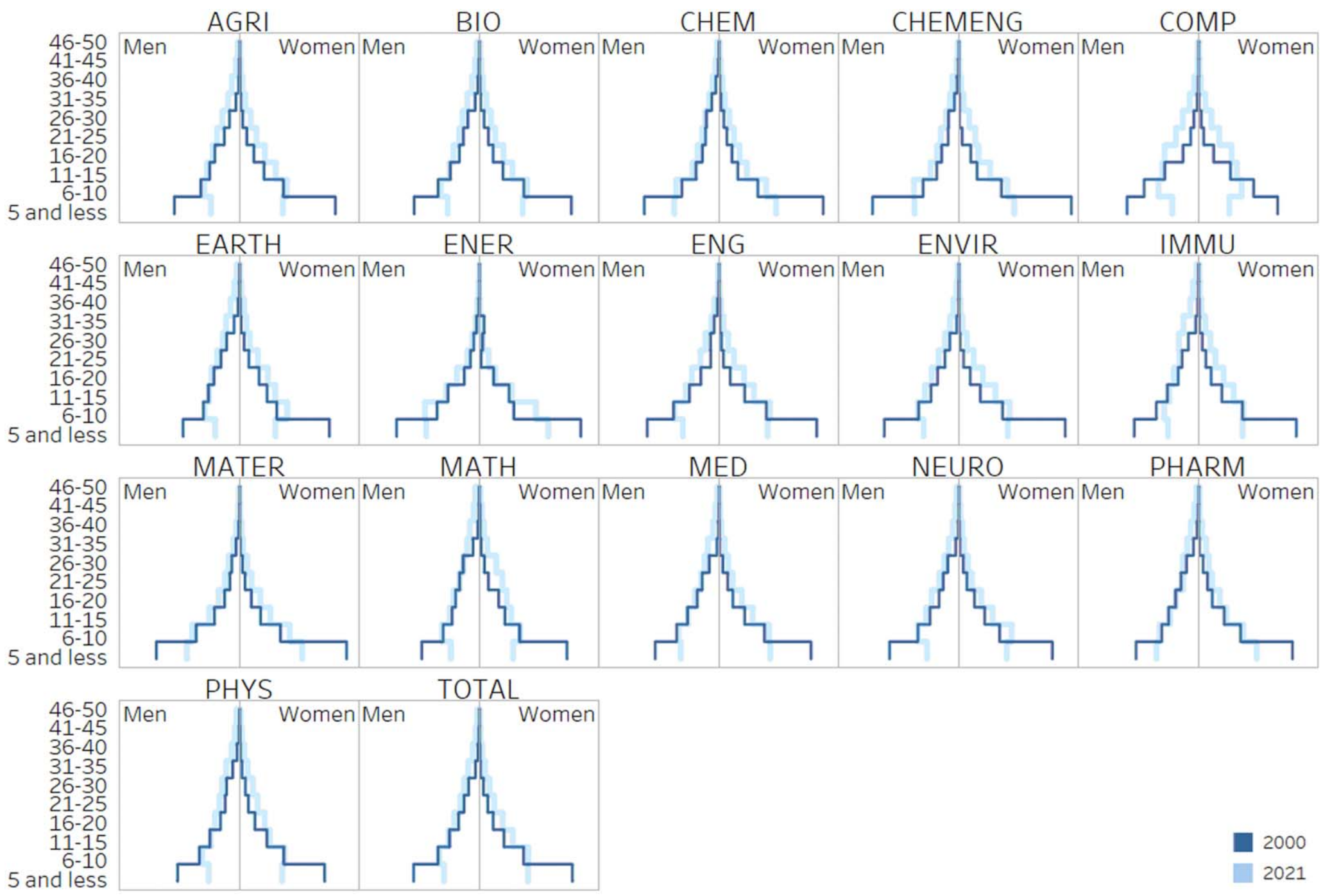

**Figure 8.** Shrinking percentages of the youngest male and female scientists among all male and female scientists over time, across all disciplines. Overview of change directions in percentages, 2000 vs. 2021: vertical approach. Distribution of nonoccasional publishing STEMM scientists by discipline, age group, and gender (column percentages: 100% vertically for all age groups combined, dark blue 2000, light blue 2021) ($N_{2021}$ = 1,502,792, $N_{2000}$ = 716,796).

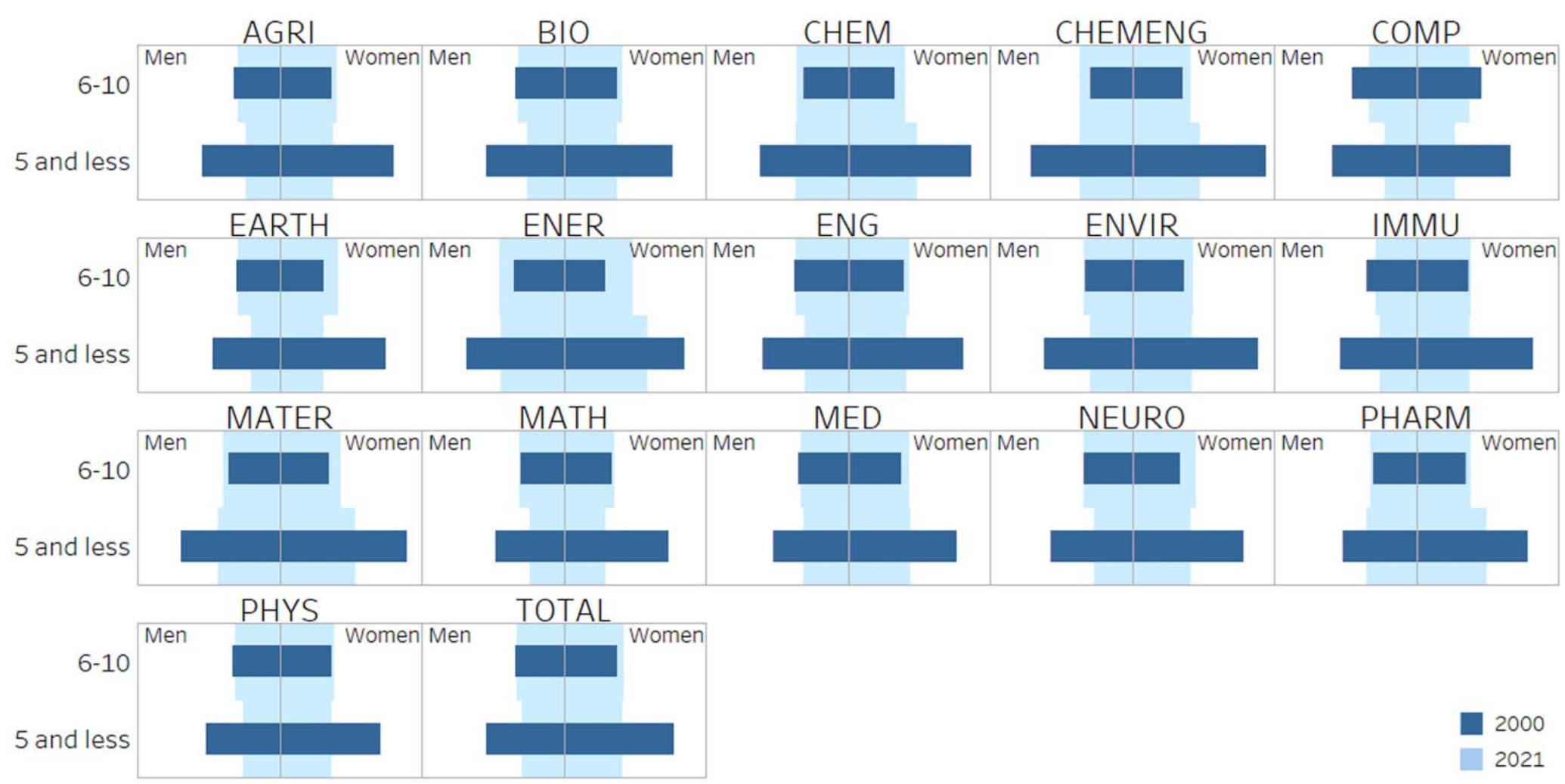

**Figure 9.** Shrinking base of young scientists, both men and women, over time. Overview of percentage change directions, 2000 vs. 2021: vertical approach. Zooming in on young scientists only (academic age 10 years or less). Distribution of young publishing nonoccasional STEMM scientists by discipline, age group, and gender, 2000 (dark blue) and 2021 (light blue) (based on column percentages) ($N_{2021}$ = 666,355, $N_{2000}$ = 437,113).

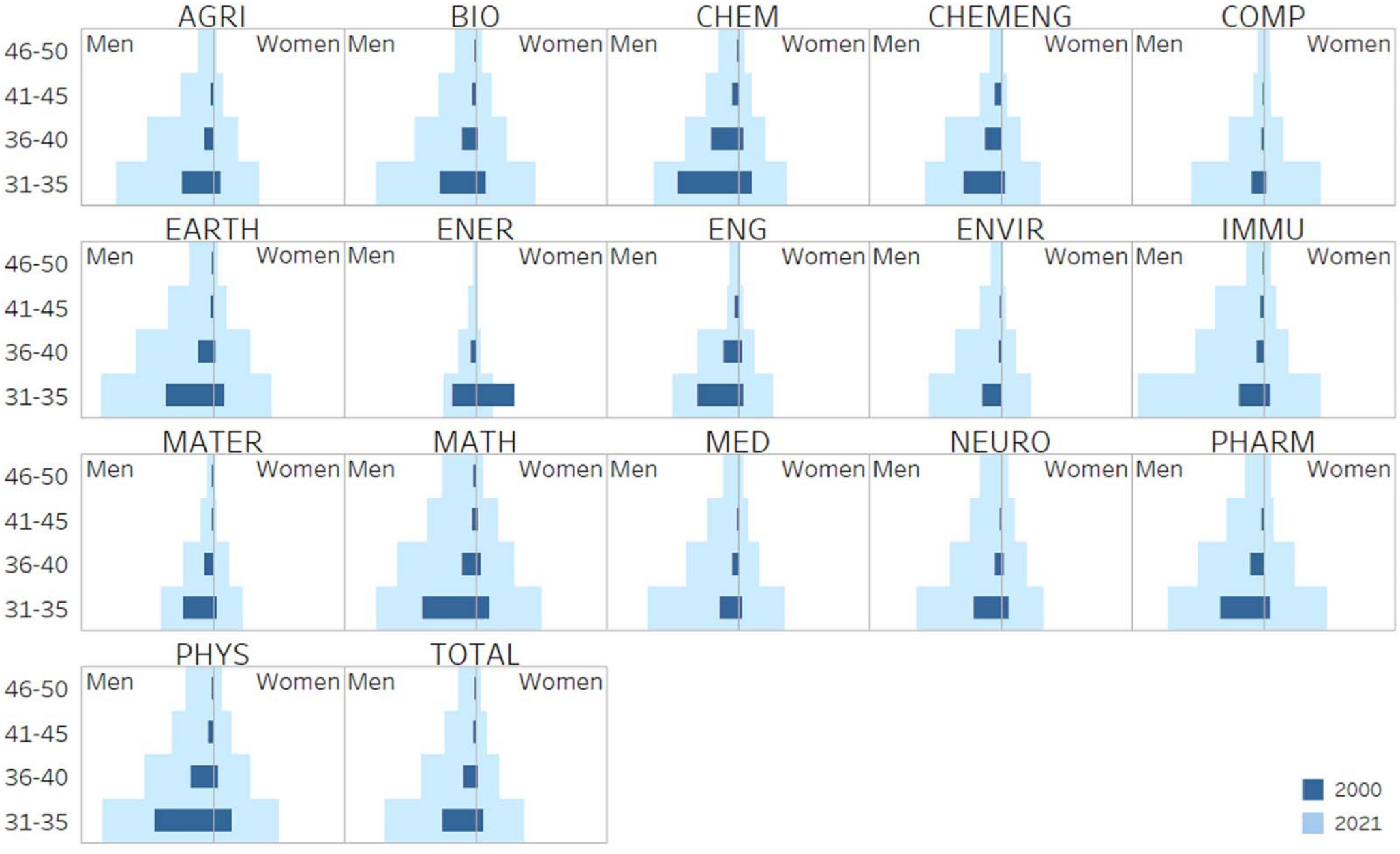

**Figure 10.** Expanding base of old scientists, both men and women, over time. Overview of change directions, 2000 vs. 2021: vertical approach. Zooming in on old scientists only: academic age of 31–50 years. Distribution of old publishing nonoccasional STEMM scientists by discipline, age group, and gender, 2000 (dark blue) and 2021 (light blue) (based on column percentages) ($N_{2021}$ = 146,090, $N_{2000}$ = 17,463).

**Table 6.** Regression model statistics: trends in the percentage of female scientists by discipline, 1990–2021

| | **Coefficient – Slope** | | | | | | **Coefficient – Intercept** | | | | **Quality measures** | |
|---|---|---|---|---|---|---|---|---|---|---|---|---|
| **Discipline** | **Value** | **Standard error** | ***t*-value** | ***p*-value** | **LB** | **UB** | **Value** | **Standard error** | ***t*-value** | ***p*-value** | **$R^2$** | **Standard error** |
| AGRI | 0.73 | 0.012 | 60.513 | < 0.0001 | 0.704 | 0.753 | 19.51 | 0.217 | 89.842 | < 0.0001 | 0.992 | 0.629 |
| BIO | 0.42 | 0.018 | 23.611 | < 0.0001 | 0.381 | 0.454 | 32.28 | 0.319 | 101.175 | < 0.0001 | 0.949 | 0.924 |
| CHEM | 0.49 | 0.021 | 22.715 | < 0.0001 | 0.443 | 0.531 | 17.79 | 0.387 | 46.011 | < 0.0001 | 0.945 | 1.120 |
| CHEMENG | 0.59 | 0.019 | 31.706 | < 0.0001 | 0.549 | 0.624 | 9.67 | 0.334 | 28.983 | < 0.0001 | 0.971 | 0.966 |
| COMP | 0.28 | 0.005 | 55.445 | < 0.0001 | 0.272 | 0.292 | 8.81 | 0.092 | 96.007 | < 0.0001 | 0.990 | 0.266 |
| EARTH | 0.56 | 0.012 | 47.946 | < 0.0001 | 0.538 | 0.586 | 12.39 | 0.211 | 58.632 | < 0.0001 | 0.987 | 0.612 |
| ENER | 0.49 | 0.014 | 34.941 | < 0.0001 | 0.466 | 0.524 | 5.64 | 0.255 | 22.089 | < 0.0001 | 0.976 | 0.739 |
| ENG | 0.27 | 0.004 | 67.841 | < 0.0001 | 0.263 | 0.279 | 7.60 | 0.072 | 105.555 | < 0.0001 | 0.994 | 0.209 |
| ENVIR | 0.81 | 0.015 | 53.321 | < 0.0001 | 0.778 | 0.840 | 17.17 | 0.274 | 62.716 | < 0.0001 | 0.990 | 0.793 |
| IMMU | 0.52 | 0.019 | 26.908 | < 0.0001 | 0.481 | 0.560 | 36.33 | 0.349 | 104.162 | < 0.0001 | 0.960 | 1.010 |
| MATER | 0.52 | 0.016 | 32.913 | < 0.0001 | 0.483 | 0.547 | 13.28 | 0.282 | 47.001 | < 0.0001 | 0.973 | 0.818 |
| MATH | 0.33 | 0.014 | 23.208 | < 0.0001 | 0.302 | 0.361 | 11.25 | 0.258 | 43.670 | < 0.0001 | 0.947 | 0.746 |
| MED | 0.71 | 0.012 | 57.674 | < 0.0001 | 0.684 | 0.734 | 26.01 | 0.222 | 117.338 | < 0.0001 | 0.991 | 0.642 |
| NEURO | 0.47 | 0.014 | 32.908 | < 0.0001 | 0.436 | 0.494 | 31.22 | 0.255 | 122.402 | < 0.0001 | 0.973 | 0.738 |
| PHARM | 0.61 | 0.020 | 30.966 | < 0.0001 | 0.568 | 0.649 | 27.98 | 0.355 | 78.940 | < 0.0001 | 0.970 | 1.026 |
| PHYS | 0.28 | 0.006 | 47.373 | < 0.0001 | 0.269 | 0.294 | 8.60 | 0.107 | 80.276 | < 0.0001 | 0.987 | 0.310 |
| **Total** | **0.55** | **0.010** | **56.971** | **< 0.0001** | **0.535** | **0.574** | **22.66** | **0.176** | **129.059** | **< 0.0001** | **0.991** | **0.508** |

#### *3.3.2. A comparative vertical view (2000 vs. 2021)*

In this section, we discuss the change in age pyramids (distributions) over 2 decades by comparing the age pyramids in 2021 and 2000. Longitudinal research measures the differences or changes in a variable between distinct periods. An age pyramid consists of paired bar graphs for men and women, with the vertical axis representing age. The 2021 age pyramids (light blue) are superimposed over the 2000 pyramids (dark blue). An age pyramid is made up of a pair of bar graphs—one for men and one for women—turned on their sides and joined, where the vertical axis corresponds to age. For each of the 10 age groups in our population, the bar coming off the axis to the right represents the share of women in that group, and the bar to the left represents the share of men (see Wachter, 2014, pp. 218–221). Both age pyramids cover a different population (there are incoming and outgoing scientists in each case); however, some of the cohorts of scientists were found to be common. The included scientists were publishing between 1970 and 2021 (for 2021 data) and 1940 and 1990.

In Figure 8, we present the percentages of male and female scientists among nonoccasional publishing authors at two points in time, disregarding the number of authors. Using the same sampling principles, this approach allows us to compare demographics at two points in time and focus on young (and old) scientists. Figure 8 displays the snapshots of 2021 and 2000 by age groups and gender, showing the distribution of male and female scientists by age group in each discipline and illustrating the dynamics of change. Although Section 3.3 uses trend analysis to demonstrate the change in female scientist percentages by discipline, this section adds age to the analysis.

In general terms, each discipline exhibits a pyramid-like demographic structure, where biological age is replaced with academic or professional age. For each discipline, the age pyramid narrows at the top and expands at the bottom, to varying degrees. The bottom represents the percentage of young scientists among all scientists, and the top signifies the percentage of older scientists. A wider bottom indicates a higher percentage of young scientists.

A common pattern emerged across all disciplines in 2021: The age pyramid's bottom (first age group, 5 years and less) was narrower compared with 2 decades earlier for both male and female scientists. The share of young female scientists among all female scientists decreased significantly compared with smaller decreases for young male scientists (see Figure 9). This decrease could also indicate that young female scientists who entered academia two decades ago remained in the system in 2021, increasing their shares in older age groups. The shrinking bottom for female scientists in 2021 compared with 2000 is also visible for all disciplines combined (Total). In terms of age structures in demographics (Rowland, 2014, pp. 98–107), the 2000 age structures can be classified as "very young" and the 2021 structures as "young" or "mature."

**Table 7.** Trends in the percentage of female scientists by discipline (slope, intercept, and speed of change), 1990–2021

| Discipline | Slope | Intercept | Time needed for a 1 p.p. change (in years) | Time needed to achieve gender parity (women 50%) in years, and the date | Time needed to achieve gender balance (women 40%) in years, and the date |
|---|---|---|---|---|---|
| ENVIR | 0.81 | 17.17 | 1.24 | 13.5 (2035) | 0 (achieved) |
| AGRI | 0.73 | 19.51 | 1.37 | 16.1 (2038) | 0 (achieved) |
| MED | 0.71 | 26.01 | 1.41 | 40.6 (2062) | 0 (achieved) |
| PHARM | 0.61 | 27.98 | 1.64 | 5.4 (2027) | 0 (achieved) |
| CHEMENG | 0.59 | 9.67 | 1.70 | 36.6 (2058) | 19.6 (2041) |
| EARTH | 0.56 | 12.39 | 1.78 | 39.4 (2061) | 21.6 (2043) |
| IMMU | 0.52 | 36.33 | 1.92 | 0 (achieved) | 0 (achieved) |
| MATER | 0.52 | 13.28 | 1.94 | 44.4 (2066) | 25.0 (2046) |
| ENER | 0.49 | 5.64 | 2.02 | 60.0 (2081) | 39.8 (2061) |
| CHEM | 0.49 | 17.79 | 2.05 | 40.6 (2062) | 20.1 (2042) |
| NEURO | 0.47 | 31.22 | 2.15 | 13.4 (2035) | 0 (achieved) |
| BIO | 0.42 | 32.28 | 2.39 | 16.1 (2038) | 0 (achieved) |
| MATH | 0.33 | 11.25 | 3.02 | 90.5 (2112) | 60.3 (2081) |
| COMP | 0.28 | 8.81 | 3.55 | 112.9 (2134) | 77.39 (2099) |
| PHYS | 0.28 | 8.60 | 3.55 | 118.5 (2140) | 83.3 (2105) |
| ENG | 0.27 | 7.60 | 3.69 | 133.5 (2155) | 96.6 (2118) |
| **Total** | **0.55** | **22.66** | **1.82** | **–** | **–** |

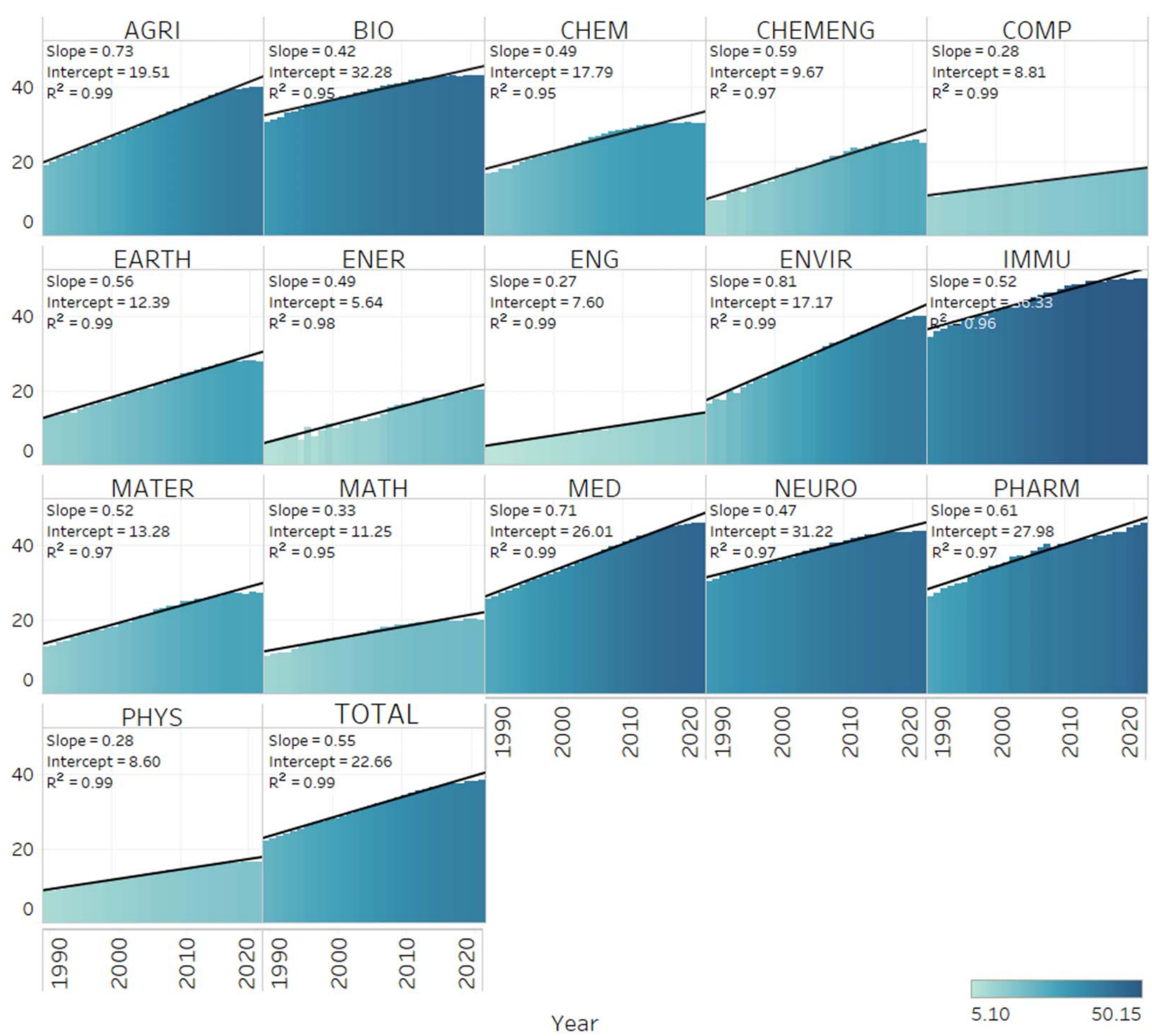

**Figure 11.** Different starting points and growth in participation of women in science over time. The trend in the percentage of female scientists by discipline, 1990–2021 ($N$ = 4,314,666).

In contrast, when comparing the shares of older male and female scientists in 2000 and 2021 within disciplines (Figure 10), the pattern is clear: The shares of both genders in the four older age groups were much higher in 2021 than in 2020. There was a higher percentage of older scientists in 2021 than in 2020 in each older category for each discipline, without exceptions, highlighting the graying of the scientific workforce.

### 3.4. Results: Trends 1990–2021, Female Scientists by Discipline

In this section, we analyze the changing participation of women in science over time to test the claim that the inflow of female scientists into science over the past 3 decades was powerfully differentiated by discipline (Figure 11).

The number of individual scientists used here to examine the trend over time was 4.3 million (61.85% male and 38.15% female, Table 1). We studied the trend of the percentage of female scientists present in global science in 1990–2021. Our analysis used a linear trend in the form of $y = at + b$. In the equation, $b$ is where the line intersected the $y$-axis and $a$ denotes the slope of the line. The slope describes how steep a line is by using a positive or negative value. The slope of *a* indicates the average change from year to year, and *b* is the intercept indicating the level of the phenomenon in the zero period (preceding the first year of analysis).

In some disciplines, women's participation in science was high with strong growth (MED and PHARM) or high with weak growth (BIO); in others, participation was low with strong growth (AGRI, CHEMENG). The Big Four, the cluster of four math-intensive disciplines, had low participation and weak growth: COMP, ENG, MATH, and PHYS. For all disciplines combined (Total), the increase was substantial, from 22.16% to 38.55%. The percentage of female scientists has been rising yearly in all disciplines, though at varying rates. MATH, COMP, PHYS, and ENG had the lowest increase, with slopes equal to or smaller than 0.33 (Table 6). All slopes were significantly positive, indicating an upward trend in female scientists' percentages across disciplines. The confidence intervals of slopes revealed specific groups' average growth rates per year. Each discipline had a different time for a one percentage point increase in female scientists' percentage. The fastest growth occurred for ENVIR (1.24 years), AGRI (1.37), and MED (1.41). Nine disciplines took slightly longer (1.64–2.39 years), and the Big Four of MATH, COMP, PHYS, and ENG took the longest, with 3.03 to 3.69 years (Table 7).

Hypothetically, under stable conditions of professional access to disciplines and current trends in women's participation in science by discipline, here based on the past 3 decades, none of which can be guaranteed in the future, male–female parity within a discipline, that is, 50% female scientists and 50% male scientists, in the four disciplines can be expected to be reached about a century from today: after 90.5 years for MATH (year 2112), 112.9 years for COMP (year 2134), 118.5 years for PHYS (year 2140), and 133.5 years for ENG (year 2155); across all other disciplines, parity can be reached between 2027 (PHARM) and 2081 (ENER). The only discipline in which gender parity has already been achieved is IMMU (see Table 7 for details). To calculate the date for gender parity for any discipline, we took the percentage points missing to reach the 50% parity level from Table 3 and multiplied the missing number of years by the time needed to reach 1 p.p. change.

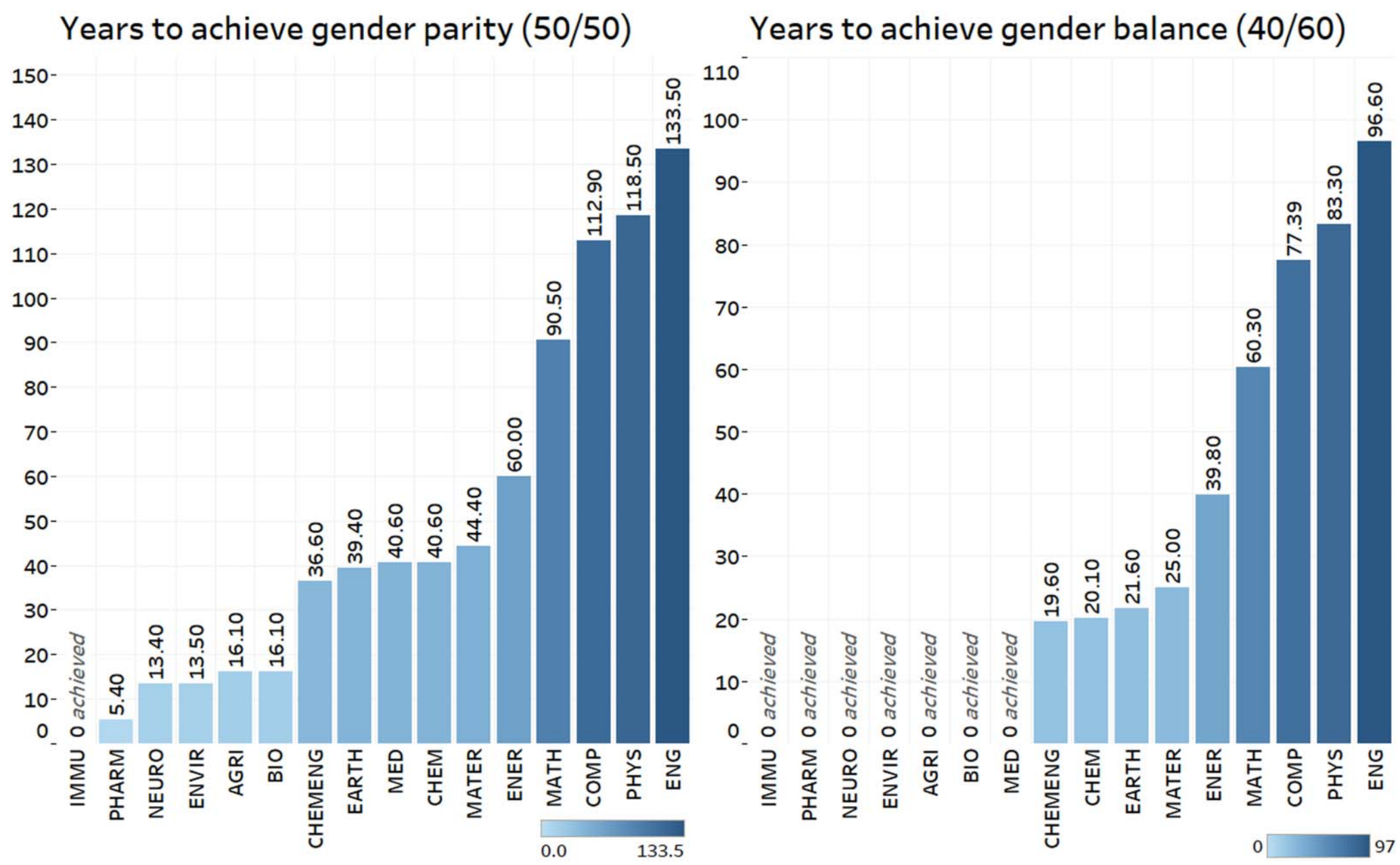

**Figure 12.** Gender parity (50/50) vs. gender balance (40/60), time needed to achieve, in years, by discipline.

Instead of gender parity (50%/50%) for all, we can focus on gender parity for the youngest generations of scientists only. And we can recalculate the results for the new age group of interest, with parity already achieved in six disciplines (e.g., AGRI, BIO, and MED; see Figure 4), and almost achieved for all disciplines combined. Instead of gender parity, we can also take an alternative approach: gender balance that refers to a presence of men and women in science that ranges between 40% and 60% of the total population (EC, 2021, p. 20). We recalculated the results for gender balance for all, with much shorter periods in which it can be achieved and with seven disciplines in which gender balance was already achieved (Table 7 and Figure 12). However, predictive analytics was outside of our scope.

## 4. SUMMARY, DISCUSSION, AND CONCLUSIONS

We have examined the changing demographics of the global scientific workforce over the past 3 decades (as defined in this research: STEMM disciplines, 38 OECD countries, nonoccasional status in science, articles indexed in the Scopus database), with special emphasis on the changing participation in the science of young male and female scientists. Our research was large scale (4.3 million scientists); generational (scientists were allocated to 10 academic age groups, with a major distinction between the young cohort, academic experience 10 or less years, and the old cohort, 31–50 years); and both cross-sectional (2021) and longitudinal (in a broader sense, the 1990 to 2021 period and 2000 vs. 2021).

We combined two approaches to comprehensively examine the four dimensions (gender, age, discipline, and time): In what we termed a *horizontal* approach, we focused on the gender distribution of scientists within the same age groups across disciplines; and in what we termed the *vertical* approach, we focused on the concentration of male and female scientists separately across age groups and within disciplines.

Our underlying methodological choice was to use individual scientists (with their attributes) rather than individual publications (with their characteristics) as a unit of analysis. We used raw microdata at an individual level from the Scopus data set because our research heavily relied on author identifiers and because Scopus provided bibliometric data with a precision of 98.1% and recall of 94.4% (Baas, Schotten et al., 2020). Our study was quantitative and exploratory in nature: Appropriately measured large scale exploratory data can set broad baseline understanding of complex issues and serve as the foundation for more specific research questions. Therefore, the present research can be complemented with further small-scale quantitative studies (based on global and national survey data) and qualitative studies based on interview and focus group methodologies (as Fox (2020) suggests in studying gender and rank). We are not aware of a similar research exercise mapping young men and women STEMM scientists across disciplines in the context of older age groups (in terms of academic or professional experience). Although statistical reports (as described in the Introduction) providing the data on men and women in science are extremely useful, they do not seem to enter the global scholarly conversation on women in science.

We did not test the various hypotheses about gender disparities in science because we have carried out an exploratory exercise; however, our findings support selected findings from the theoretical background discussed in the Introduction section in general terms. Attrition levels for female scientists are high ("leaky pipeline" theory), there are clearly disciplines which—for some reason—are not welcoming to women ("chilly climate" hypothesis), and in which the generational structure of the scientific workforce is not changing ("self-selection" hypothesis; see the Big Four disciplines and its stable age and gender distribution over time).

The scientific workforce has been changing in terms of its gender and age composition, with different intensities in different disciplines. These changes have been ongoing and global in nature. Among the 16 STEMM disciplines, most were currently numerically dominated by men, but some were already dominated by women, and the change processes seemed to be fast in some and slow in others. A surprising finding, even in the context of the COVID-19 pandemic, was the pivotal role of medical research for the global scientific workforce, especially for female scientists: Almost half of all scientists (45.98%) were defined in our methodology as doing medical research (a dominating discipline, based on all cited references from lifetime publications for each scientist). The concentration of female scientists was steep across disciplines: More than half (55.02%) were located in MED and 1 in 7 (15.91%) in BIO. Consequently, about 70% (70.93%) of all female scientists globally, across all STEMM science sectors, were concentrated just in these two disciplines.

The traditional narratives about some STEMM disciplines being much more heavily male dominated than others have been confirmed: Women's participation in COMP, ENG, MATH, and PHYS was very low (and smaller than 20% in 2021). In most disciplines in 2021, the share of female scientists in each successive younger cohort was higher (and it was usually the highest for the youngest cohort: scientists with 5 or less years of academic experience); for COMP, ENG, MATH, and PHYS, however, the principle did not hold, with very small intracohort differences (Figure 3).

Our trend analysis of the 1990–2021 period showed that the participation of female scientists in global science increased across all disciplines, albeit with different starting points in 1990 and different intensities, following an array of past research on "women in science." For the least increasing trends, the increase in the percentage of female scientists by one percentage point took 3.03 years for MATH, 3.55 for COMP and PHYS, and 3.69 years for ENG. Hypothetically, male–female parity within a discipline (50% female scientists, 50% male scientists) in the four disciplines can be expected to be reached about a century from today: for MATH in 2112, for COMP in 2134, for PHYS in 2140, and for ENG in 2155; across all other disciplines, parity can be reached between 2027 (PHARM) and 2081 (ENER). In a less restricted approach, gender balance (40% female scientists, 60% male scientists) has already been achieved in seven disciplines; see Figure 12 for details.

However, from an age-disaggregated perspective, in six out of 16 disciplines, there were already more youngest female than male scientists (IMMU, PHARM, NEURO, MED, AGRI, BIO), and the discipline most open to female scientists has been IMMU (59.04%). Interestingly, more than eight out of 10 STEMM female scientists globally worked in these six disciplines (82.90%). Across all STEMM disciplines combined, the majority of women currently involved in publishing articles were young women (with 10 years of academic experience or less).

Most interestingly, there was a higher concentration of young women than young men across all STEMM disciplines, and there was a higher concentration of old men than old women across all disciplines. For every discipline, the share of young female scientists among all female scientists within a discipline was higher than the share of young male scientists among all male scientists. For every discipline, the share of old male scientists among all male scientists within a discipline was substantially higher than the share of old female scientists among all female scientists. The patterns are clear: For all STEMM disciplines, female scientists were generally younger and male scientists generally older.

Moving from standard data (of the OECD, UNESCO and Eurostat type) to gender-disaggregated data for particular age groups, we begin to understand what the global isolation of female scientists in such disciplines as MATH, PHYS, and ENG means. In these disciplines, in 2021, the share of old female scientists was about 10% or less (the difference in numbers by gender was about 10-fold or higher, e.g., ENG, MATH, and PHYS: 6.31%, 11.09%, and 9.21%, respectively). In older generations, female scientists were isolated individuals among their similar-age male colleagues. The numbers show more than percentages (Table 5): For instance, in the 36–40 academic age group, there were 84 female scientists globally working alongside 1,466 male scientists in ENG and 396 female scientists working alongside 3,726 male scientists in PHYS.

However, the context of changing times is important: For the same three disciplines of ENG, MATH, and PHYS, the isolation of young female scientists powerfully decreased, from a 10-times difference for older cohorts to five times difference for young cohorts (i.e., to 16.47% for ENG, 22.04% for MATH, and 20.23% for PHYS). In these three male-dominated disciplines in 2021, female scientists in young cohorts were at least twice as present as female scientists in older cohorts (on the role of gender team composition in science, see Fox and Mohapatra (2007)).

The change in gender participation in science has been gradual and the pattern unambiguous: Across all STEMM disciplines, both those heavily male dominated and those closest to gender parity, the younger generations have generally always had more female scientists and higher percentages than older generations. Female scientists were more present in numbers and more present in percentages going down the 10 age groups and when moving from the cohort of old scientists to that of young scientists. From a longitudinal perspective, in a broader sense, for all disciplines, the share of scientists in the youngest age group in 2000 was higher than in 2021 for both male and female scientists. There was a shrinking base of young scientists, both men and women, and there was an expanding base of old scientists, both men and women.

Most limitations of bibliometric data sets have been widely discussed (English language and STEMM focus, Anglo-Saxon bias, articles only, etc.; see Sugimoto and Larivière (2018, pp. 38–44) on “cultural biases of data sources”). However, our use of a bibliometric data set to define the individual attributes of the global scientific workforce requires a brief discussion of new limitations, as follows:

1. Gender determination: A binary approach was used with different coverage for different countries, as algorithms used by Scopus (and other gender-determining tools such as Genderize.io or Gender Guesser; see Halevi [2019, p. 566] and Mihaljević and Santamaría [2020, pp. 1477–1478]) work much better for some rather than for other countries; all gender-unknown cases were removed from our analysis.
2. Discipline determination: A commercial academic journal classification was used as a proxy for the richness of nationally defined academic disciplines and lifetime Scopus-indexed publication history, with lifetime cited references being used to determine a single attribute of discipline (a single dominant value, possibly suppressing the changes between disciplines over time).
3. Determining the country of affiliation: A single dominant value, possibly suppressing individual lifetime migration histories.
4. Determining scientists’ nonoccasional status: The threshold of three articles as an entry condition for inclusion in the population was arbitrary, underplaying the role of scientists in very early stages of academic careers; a higher threshold would decrease the population, and a lower one would increase it.
5. Determining academic age: Although the correlation between biological age and academic age in the STEMM disciplines was high (and possibly higher than 0.9, as we have shown for a sample of 20,000 Polish scientists with doctorates; Kwiek & Roszka, 2022a), the first publications in individual lifetime publication histories may appear in different moments of academic lives in different disciplines; additionally, publishing patterns clearly change over time; that is, scientists tend to start publishing earlier in their careers today than before.

Another takeaway is that there were clear differences between national-level studies, especially when bibliometric data were merged with administrative and biographical data, and a global study of the academic workforce and careers. In short, national studies can use commercial and noncommercial data sets available for a few countries only (e.g., the United States, Norway, Poland, and Italy: see Abramo, Aksnes, & D'Angelo, 2021; Abramo, D'Angelo, & Murgia, 2016; Savage & Olejniczak, 2021), which may include globally directly unavailable biographical information, such as gender, date of birth, dates of PhD and other degrees and ranks, national discipline classifications, and full employment history. In our two recent longitudinal (in a narrow sense) studies of changing productivity classes of 2,326 full professors over 20–40 years of their careers (Kwiek & Roszka, 2023a) and of the impact of early and late, as well as fast and slow promotions on productivity on a sample of 16,000 STEMM university professors (Kwiek & Roszka, 2023b), our data set of about a million Polish Scopus-indexed publications from the past 50 years was enriched with full biographical and administrative data from a registry of 100,000 Polish scientists.

In global studies—as opposed to national studies—biological age needs to be examined through a proxy of academic or professional age, gender needs to be inferred with probability thresholds, academic ranks should be used through a proxy of career length from the first publication, and national prestige ranks should be used through a proxy of global rankings. All scientists registered nationally must be replaced in global studies with publishing-only scientists, with Scopus- (or WoS-) indexed publications. Real scientists with national identification numbers available in national databases need to be replaced with Scopus Author IDs, and near-perfect administrative and biographical data need to be replaced with either inferred data or proxies. However, global exploratory research, provisionally mapping the terrain and testing the best tools and methodologies, is interesting in its generality before more sophisticated analyses arise. The world of Scopus authors (and their Scopus-indexed publication) is not the real world of science—but it may be a useful proxy for it.

The scholarly and policy implications of the present research are manifold. In scholarly terms, we make the first attempt to define the scientific community globally through attributes so far understudied on a large scale. The mapping of changing gender and age distribution of scientists globally over time, as well as a glimpse of the global scientific workforce today, opens science (and academic) profession studies to more detailed questions. The scientific workforce is often discussed in two policy contexts: the aging and accompanying problems for higher education and innovation systems and access to the science profession of young scientists. Our methodological approach and findings can be useful in examining the complex policy issue of entering and leaving the science profession, with the accompanying questions about changing productivity over scientists' life cycles, aging, and changing publishing and collaboration patterns, and so forth (especially in the academic sector).

Our research can be useful for policymakers, administrators, and large grant-making organizations in showing where the scientific workforce has been focusing their research efforts, how large segments of academics are involved in studies in particular disciplines, and where male and female scientists are disciplinary located. Our mapping of substantial gender differences between the various STEMM disciplines (and especially between ENG, COMP, MATH, and PHYS versus all others) may provide new empirical grounds that are useful in discussing women's participation in science and its discipline-based social, institutional, and political impediments.

## ACKNOWLEDGMENTS

Marek Kwiek is grateful for the comments from the hosts and audiences of the four invited seminars about this strand of research: the RISIS (Research Infrastructure for Science and Innovation Policy Studies) Research Seminar hosted by Benedetto Lepori, University of Lugano, Switzerland (May 11, 2022); the METRICS (Meta-Research Innovation Center) seminar hosted by John Ioannidis, Stanford University (February 23, 2023); the DZHW (German Center for Higher Education Research and Science Studies, Berlin) seminar hosted by Torger Möller (January 10, 2023), and the CGHE (Center for Global Higher Education) seminar hosted by Simon Marginson, University of Oxford (April 4, 2023). We gratefully acknowledge the assistance of the International Center for the Studies of Research (ICSR) Lab and Kristy James, Senior Data Scientist. We also want to thank Dr Wojciech Roszka from the CPPS Poznan Team for many fruitful discussions. We are also very grateful to the three anonymous reviewers for their penetrating comments.

## AUTHOR CONTRIBUTIONS

Marek Kwiek: Conceptualization, Data curation, Formal analysis, Investigation, Methodology, Resources, Software, Validation, Writing—original draft, Writing—review & editing. Lukasz Szymula: Conceptualization, Data curation, Formal analysis, Investigation, Methodology, Software, Validation, Visualization, Writing—original draft, Writing—review & editing.

## COMPETING INTERESTS

The authors have no competing interests.

## FUNDING INFORMATION

Marek Kwiek gratefully acknowledges the support provided by the MeiN NDS grant no. NdS/529032/2021/2021. Lukasz Szymula is grateful for the support of his doctoral studies provided by the NCN grant 2019/35/0/HS6/02591.

## DATA AVAILABILITY

We used data from Scopus, a proprietary scientometric database. For legal reasons, data from Scopus received through collaboration with the ICSR Lab cannot be made openly available.

## REFERENCES

Abramo, G., Aksnes, D. W., & D'Angelo, C. A. (2021). Gender differences in research performance within and between countries: Italy vs Norway. *Journal of Informetrics*, *15*(2), 101144. https://doi.org/10.1016/j.joi.2021.101144

Abramo, G., D'Angelo, C. A., & Murgia, G. (2016). The combined effects of age and seniority on research performance of full professors. *Science and Public Policy*, *43*(3), 301–319. https://doi.org/10.1093/scipol/scv037

Arimoto, A., Cummings, W. K., Huang, F., & Shin, J. C. (2015). *The changing academic profession in Japan*. Cham: Springer. https://doi.org/10.1007/978-3-319-09468-7

Baas, J., Schotten, M., Plume, A., Côté, G., & Karimi, R. (2020). Scopus as a curated, high-quality bibliometric data source for academic research in quantitative science studies. *Quantitative Science Studies*, *1*(1), 377–386. https://doi.org/10.1162/qss_a_00019

Boekhout, H., van der Weijden, I., & Waltman, L. (2021). Gender differences in scientific careers: A large-scale bibliometric analysis. *arXiv*. https://doi.org/10.48550/arXiv.2106.12624

Boothby, C., Milojevic, S., Larivière, V., Radicchi, F., & Sugimoto, C. (2022). Consistent churn of early career researchers: An analysis of turnover and replacement in the scientific workforce. Preprint. https://doi.org/10.31219/osf.io/hdny6

Börner, K. (2010). *Atlas of science: Visualizing what we know*. Cambridge, MA: MIT Press.

Britton, D. M. (2017). Beyond the chilly climate: The salience of gender in women's academic careers. *Gender & Society*, *31*(1), 5–27. https://doi.org/10.1177/0891243216681494
Calikoglu, A., Jones, G. A., & Kim, Y. (2023). *Internationalization and the academic profession. Comparative perspectives*. Cham: Springer. https://doi.org/10.1007/978-3-031-26995-0
Carvalho, T. (2017). The study of the academic profession—Contributions from and to the sociology of professions. In J. Huisman & M. Tight (Eds.), *Theory and method in higher education research* (pp. 59–76). Emerald. https://doi.org/10.1108/S2056-375220170000003004
Cornelius, R., Constantinople, A., & Gray, J. (1988). The chilly climate: Fact or artifact? *Journal of Higher Education*, *59*(5), 527–550. https://doi.org/10.2307/1981702
Cummings, W. K., & Finkelstein, M. J. (2012). *Scholars in the changing American academy: New contexts, new rules and new roles*. Dordrecht: Springer. https://doi.org/10.1007/978-94-007-2730-4
Cummings, W. K., & Teichler, U. (Eds.). (2015). *The relevance of academic work in comparative perspective*. Cham: Springer. https://doi.org/10.1007/978-3-319-11767-6
Dusdal, J., & Powell, J. J. W. (2021). Benefits, motivations, and challenges of international collaborative research: A sociology of science case study. *Science and Public Policy*, *48*(1), 235–245. https://doi.org/10.1093/scipol/scab010
Elsevier. (2015). *Mapping gender in the german research arena*. Elsevier.
Elsevier. (2017). *Gender in the global research landscape: Analysis of research performance through a gender lens across 20 years, 12 geographies, and 27 subject areas*. Elsevier.
Elsevier. (2020). *The researcher journey through a gender lens*. Elsevier.
European Commission. (2021). *She Figures*. European Commission.
Fox, M. F. (2020). Gender, science, and academic rank: Key issues and approaches. *Quantitative Science Studies*, *1*(3), 1001–1006. https://doi.org/10.1162/qss_a_00057
Fox, M. F., & Mohapatra, S. (2007). Social-organizational characteristics of work and publication productivity among academic scientists in doctoral-granting departments. *Journal of Higher Education*, *78*(5), 542–571. https://doi.org/10.1353/jhe.2007.0032
Fox, M. F., & Nikivincze, I. (2021). Being highly prolific in academic science: Characteristics of individuals and their departments. *Higher Education*, *81*, 1237–1255. https://doi.org/10.1007/s10734-020-00609-z
Fumasoli, T., Goastellec, G., & Kehm, B. M. (Eds.). (2015). *Academic work and careers in Europe: Trends, challenges, perspectives*. Cham: Springer. https://doi.org/10.1007/978-3-319-10720-2
Halevi, G. (2019). Bibliometric studies on gender disparities in science. In W. Glänzel, H. F. Moed, U. Schmoch, & M. Thelwall (Eds.), *Springer handbook of science and technology indicators* (pp. 563–580). Cham: Springer. https://doi.org/10.1007/978-3-030-02511-3_21
Hall, R., & Sandler, B. R. (1982). *The classroom climate: A chilly one for women?* Washington, DC: Association of American Colleges.
Holmes, D. E. (2017). *Big data: A very short introduction*. Oxford University Press. https://doi.org/10.1093/actrade/9780198779575.001.0001
Huang, F., Finkelstein, M., & Rostan, M. (2014). *The internationalization of the academy: Changes, realities and prospects*. Dordrecht: Springer. https://doi.org/10.1007/978-94-007-7278-6
Huang, J., Gates, A. J., Sinatra, R., & Barabási, A.-L. (2020). Historical comparison of gender inequality in scientific careers across countries and disciplines. *Proceedings of the National Academy of Sciences*, *117*(9), 4609–4616. https://doi.org/10.1073/pnas.1914221117, PubMed: 32071248
Hyde, J. S., Fennema, E., Ryan, M., Frost, L. A., & Hopp, C. (1990). Gender comparisons of mathematics attitudes and affect: A meta-analysis. *Psychology of Women Quarterly*, *14*(3), 299–324. https://doi.org/10.1111/j.1471-6402.1990.tb00022.x
Ioannidis, J. P. A., Boyack, K. W., & Klavans, R. (2014). Estimates of the continuously publishing core in the scientific workforce. *PLOS ONE*, *9*(7), e101698. https://doi.org/10.1371/journal.pone.0101698, PubMed: 25007173
Kanter, R. M. (1977). Some effects of proportions on group life: Skewed sex ratios and responses to token women. *American Journal of Sociology*, *82*(5), 965–990. https://doi.org/10.1086/226425
King, M. M., Bergstrom, C. T., Correll, S. J., Jacquet, J., & West, J. D. (2017). Men set their own cites high: Gender and self-citation across fields and over time. *Socius*, *3*. https://doi.org/10.1177/2378023117738903
Kwiek, M. (2016). The European research elite: A cross-national study of highly productive academics across 11 European systems. *Higher Education*, *71*(3), 379–397. https://doi.org/10.1007/s10734-015-9910-x
Kwiek, M. (2018). High research productivity in vertically undifferentiated higher education systems: Who are the top performers? *Scientometrics*, *115*(1), 415–462. https://doi.org/10.1007/s11192-018-2644-7, PubMed: 29527074
Kwiek, M. (2019). *Changing European academics: A comparative study of social stratification, work patterns and research productivity*. London: Routledge. https://doi.org/10.4324/9781351182041
Kwiek, M. (2020). Internationalists and locals: International research collaboration in a resource-poor system. *Scientometrics*, *124*, 57–105. https://doi.org/10.1007/s11192-020-03460-2
Kwiek, M., & Roszka, W. (2021a). Gender-based homophily in research: A large-scale study of man-woman collaboration. *Journal of Informetrics*, *15*(3), 1–38. https://doi.org/10.1016/j.joi.2021.101171
Kwiek, M., & Roszka, W. (2021b). Gender disparities in international research collaboration: A large-scale bibliometric study of 25,000 university professors. *Journal of Economic Surveys*, *35*(5), 1344–1380. https://doi.org/10.1111/joes.12395
Kwiek, M., & Roszka, W. (2022a). Academic vs. biological age in research on academic careers: A large-scale study with implications for scientifically developing systems. *Scientometrics*, *127*, 3543–3575. https://doi.org/10.1007/s11192-022-04363-0
Kwiek, M., & Roszka, W. (2022b). Are female scientists less inclined to publish alone? The gender solo research gap. *Scientometrics*, *127*, 1697–1735. https://doi.org/10.1007/s11192-022-04308-7
Kwiek, M., & Roszka, W. (2023a). Once highly productive, forever highly productive? Full professors' research productivity from a longitudinal perspective. *Higher Education*. https://doi.org/10.1007/s10734-023-01022-y
Kwiek, M., & Roszka, W. (2023b). The young and the old, the fast and the slow: A large-scale study of productivity classes and rank advancement. *Studies in Higher Education*. https://doi.org/10.1080/03075079.2023.2288172
Larivière, V., Ni, C., Gingras, Y., Cronin, B., & Sugimoto, C. R. (2013). Global gender disparities in science. *Nature*, *504*(7479), 211–213. https://doi.org/10.1038/504211a, PubMed: 24350369
Maranto, C. L., & Griffin, A. E. (2011). The antecedents of a 'chilly climate' for women faculty in higher education. *Human*

*Relations*, *64*(2), 139–159. https://doi.org/10.1177/0018726710377932

Menard, S. (2002). *Longitudinal research*. Thousand Oaks, CA: Sage. https://doi.org/10.4135/9781412984867

Mihaljević, H., & Santamaría, L. (2020). Authorship in top-ranked mathematical and physical journals: Role of gender on self-perceptions and bibliographic evidence. *Quantitative Science Studies*, *1*(4), 1468–1492. https://doi.org/10.1162/qss_a_00090

Morris, L. K., & Daniel, L. G. (2008). Perceptions of a chilly climate: Differences in traditional and non-traditional majors for women. *Research in Higher Education*, *49*, 256–273. https://doi.org/10.1007/s11162-007-9078-z

Morrison, A. M., White, R. P., & Van Velsor, E. (1987). *Breaking the glass ceiling: Can women reach the top of America's largest corporations?* Addison-Wesley.

Nielsen, M. W., & Andersen, J. P. (2021). Global citation inequality is on the rise. *Proceedings of the National Academy of Sciences*, *118*(7), e2012208118. https://doi.org/10.1073/pnas.2012208118, PubMed: 33558230

NSF. (2023). *Diversity and STEM: Women, minorities, and persons with disabilities*. National Science Foundation.

Nygaard, L. P., Piro, F., & Aksnes, D. (2022). Gendering excellence through research productivity indicators. *Gender and Education*, *34*(6), 690–704. https://doi.org/10.1080/09540253.2022.2041186

Robinson-Garcia, N., Costas, R., Sugimoto, C. R., Larivière, V., & Nane, G. F. (2020). Task specialization across research careers. *eLife*, *9*, e60586. https://doi.org/10.7554/eLife.60586, PubMed: 33112232

Rowland, D. T. (2014). *Demographic methods and concepts*. Oxford: Oxford University Press.

Ruspini, E. (1999). Longitudinal research and the analysis of social change. *Quality and Quantity*, *33*(3), 219–227. https://doi.org/10.1023/A:1004692619235

Salganik, M. J. (2018). *Bit by bit: Social research in a digital age*. Princeton, NJ: Princeton University Press.

Savage, W. E., & Olejniczak, A. J. (2021). Do senior faculty members produce fewer research publications than their younger colleagues? Evidence from Ph.D. granting institutions in the United States. *Scientometrics*, *126*, 4659–4686. https://doi.org/10.1007/s11192-021-03957-4

Selwyn, N. (2019). *What is digital sociology?* Cambridge: Polity Press.

Sexton, K. W., Hocking, K. M., Wise, E., Osgood, M. J., Cheung-Flynn, J., ... Brophy, C. M. (2012). Women in academic surgery: The pipeline is busted. *Journal of Surgical Education*, *69*(1), 84–90. https://doi.org/10.1016/j.jsurg.2011.07.008, PubMed: 22208838

Shaw, A. K., & Stanton, D. E. (2012). Leaks in the pipeline: Separating demographic inertia from ongoing gender differences in academia. *Proceedings. Biological Sciences*, *279*(1743), 3736–41. https://doi.org/10.1098/rspb.2012.0822, PubMed: 22719028

Sheltzer, J. M., & Smith, J. C. (2014). Elite male faculty in the life sciences employ fewer women. *Proceedings of the National Academy of Sciences of the United States of America*, *111*(28), 10107–10112. https://doi.org/10.1073/pnas.1403334111, PubMed: 24982167

Sugimoto, C., & Larivière, V. (2018). *Measuring research: What everyone needs to know*. Oxford: Oxford University Press. https://doi.org/10.1093/wentk/9780190640118.001.0001

Sugimoto, C., & Larivière, V. (2023). *Equity for women in science: Dismantling systemic barriers to advancement*. Cambridge, MA: Harvard University Press.

Tang, J. (1997). The glass ceiling in science and engineering. *Journal of Socio-Economics*, *26*(4), 383–406. https://doi.org/10.1016/S1053-5357(97)90003-2

Tang, L., & Horta, H. (2023). Supporting academic women's careers: Male and female academics' perspectives at a Chinese research university. *Minerva*. https://doi.org/10.1007/s11024-023-09506-y

Teichler, U., & Cummings, W. K. (Eds.). (2015). *Forming, recruiting, and managing the academic profession*. Cham: Springer. https://doi.org/10.1007/978-3-319-16080-1

Wachter, K. W. (2014). *Essential demographic methods*. Cambridge, MA: Harvard University Press. https://doi.org/10.4159/9780674369757

Way, S. F., Morgan, A. C., Clauset, A., & Larremore, D. B. (2017). The misleading narrative of the canonical faculty productivity trajectory. *Proceedings of the National Academy of Sciences*, *114*(44), E9216–E9223. https://doi.org/10.1073/pnas.1702121114, PubMed: 29042510

Whitt, E. J., Nora, A., Edison, M., Terenzini, P. T., & Pascarella, E. T. (1999). Women's perceptions of a "chilly climate" and cognitive outcomes in college: Additional evidence. *Journal of College Student Development*, *40*, 163–177.

Wolfinger, N. H., Mason, M. A., & Goulden, M. (2008). Problems in the pipeline: Gender, marriage, and fertility in the ivory tower. *Journal of Higher Education*, *79*(4), 388–405. https://doi.org/10.1080/00221546.2008.11772108

Zhang, S., Wapman, K. H., Larremore, D. B., & Clauset, A. (2022). Labor advantages drive the greater productivity of faculty at elite universities. *Science Advances*, *8*(46), eabq7056. https://doi.org/10.1126/sciadv.abq7056, PubMed: 36399560